\documentclass[12pt]{iopart}
\usepackage[numbers,square,sort&compress]{natbib}
\usepackage{graphicx}
\usepackage{siunitx}
\usepackage[shortlabels]{enumitem}
\usepackage[colorlinks, citecolor = blue, urlcolor = blue]{hyperref}
\usepackage{amssymb}
\usepackage{longtable}
\usepackage[T1]{fontenc}
\usepackage[utf8]{inputenc}
\usepackage{xcolor}

\begin{document}

\title{Partial ionisation cross sections for the binary-encounter Bethe model.}

\author{Anthony Jeseněk\dag, Alejandro Luque\dag, Nikolai Lehtinen\ddag}

\address{\dag Instituto de Astrof\'{i}sica de Andaluc\'{i}a, Granada, 18008, Espa\~{n}a}
\address{\ddag Birkeland Center for Space Science, University of Bergen}
\ead{anthony.schmalzried@mailfence.com}
\ead{aluque@iaa.es}
\ead{nikolai.lehtinen@uib.no}
\vspace{10pt}
\begin{indented}
\item[]July 2025
\end{indented}

\begin{abstract}
The original binary-encounter Bethe model of \citet{Kim1994} has proven to be an accurate analytical representation of total impact ionisation cross sections of electrons colliding with atoms and molecules. It is based on a decomposition into partial ionisation cross sections from electrons in bound orbitals. Despite the model's accuracy for \emph{total} ionisation, its individual \emph{partial} cross sections for ionisation rely on thresholds calculated theoretically which systematically overestimate the experimental orbital binding energies. Here, we examine the BEB model's performance when based on experimental ionisation thresholds. The resulting partial cross sections of the various final (excited) ionic states produced could help to prefigure subsequent optical radiations and non-radiative transitions in models of plasma physics. 
\end{abstract}

%
%
%
%
%

\section{Introduction}

Electron impact ionisation on atmospheric molecules is an important process in plasmas where the average kinetic energy of the electrons is near or above the ionisation threshold of the medium species. Analytical representations of cross sections are an economical and efficient way of modelling impact ionisation. There exist many semi-empirical or empirical formulae that accurately represent total ionisation cross sections \citep{Lotz1967,Hwang1996,Deutsch2004}, reviewed in \citet{EII-2}. Among these, the binary-encounter Bether (BEB) model gives remarkable results while requiring as external input only values of the electron kinetic $U$ and binding $B$ energies in occupied atomic or molecular orbitals. With theoretically
calculated values of $U$ and $B$, except for the \emph{lowest} ionisation threshold determined experimentally, the BEB was successfully applied for modelling total ionisation cross sections from atoms \citep{Kim1994,Kim2002}, diatomic molecules \citep{Hwang1996,Huber2019} and other polyatomic atmospheric molecules \citep{Kim1997}. 

The \emph{total} ionisation cross section $\sigma_\mathrm{ion}$ is the sum of all \emph{partial} cross sections from ionisation processes in which the ion is left in the ground or an excited state. When interested in tracking these excited ionic species or accurately modelling electron energy losses in ionisation, it is important to distinguish different partial ionisation channels. In principle, the BEB model already gives a decomposition of these partial processes according to the theoretical orbital structure of the target atom or molecule. In practice, however, these partial processes may \emph{not} be thus modelled, as they do not reflect the experimental \emph{threshold} nor the actual ionisation process determined by the transition from the ground neutral state to the final (excited) ionic state. 

In this paper, we discuss how can the BEB model be adapted to model partial processes of ionisation useful in Monte Carlo \cite{Kunhardt1986,Celestin2011,Koehn2014,Chanrion2010,Schmalzried2023} or kinetic \citep{Ridenti2015,Hagelaar2005,Simek2018} modelling of electron swarms in gaseous media. We begin in the next section (2) by reminding the reader about the BEB model and restating its fundamental assumptions. For each assumption, we also present its drawbacks. Then, in section 3, we introduce another perspective of the BEB, not based on orbital structure but on ionisation channels and we argue why we think it is a more appropriate application. This new perspective, we illustrate in section (4), where we apply the BEB model to the impact ionisation of electrons on simple atomic and molecular gas targets relevant for atmospheric studies. At the end, we present a conclusion (5) and an outlook for the BEB model for partial ionisation. 

\section{The orbital binary-encounter Bethe model}

The binary-encounter Bethe model is based on a Born approximation combining long-range dipole interaction and terms from the Mott cross section between two electrons for close-range binary encounters. As we are interested here to represent partial cross sections at energies of at most a few hundred eVs, we present the non-relativistic formula for the sake of simplicity. For an incident electron at a kinetic energy $\varepsilon$, the partial ionisation cross section at a threshold $I_j$ from an orbital $j$ hosting $N_j$ electrons is given by \citep[eq.(11)]{Kim2000}:

\begin{equation}
	\sigma_j(\varepsilon) = \frac{4\pi a_0^2 N_j}{\varepsilon + K_j} \frac{\text{Ryd}^2}{I_j} \left[ \frac{1}{2} \ln(\varepsilon/I_j)\left(1-\left(\frac{I_j}{\varepsilon}\right)^2\right) + 1 - \frac{I_j}{\varepsilon} - \frac{\ln(\varepsilon/I_j)}{1 + \varepsilon/I_j}  \right] \; , \label{eq:beb}
\end{equation}
with the atomic Bohr radius $a_0\approx \qty{5.291e-11}{m}$ and Rydberg energy $\text{Ryd}\approx \qty{13.602}{eV}$. In addition to the assumptions underlying the Bethe approximation, the BEB formula (\ref{eq:beb}) assumes that:
\begin{enumerate}[a)]
	\item The parameter $K_j$ represents the gain in kinetic energy of the incident electron at the place where ionisation takes place;
	\item The dipole oscillator strength density, depending on the secondary electron kinetic energy $w$, follows a simple analytical form (\ref{eq:df/dw}) $\propto 1/(w+I_j)^2$;
	\item The Bethe sum of the dipole oscillator strength is fully due to ionisation, i.e. the contribution of all discrete optically allowed excitations is accounted in the magnitude of the BEB model \citep[][$Q=1$ assumption on p.3959 at the end of §IV.]{Kim1994};
	\item The values of $K_j$ and $I_j$ can be obtained from the kinetic $U_j$ and binding $B_j$ energies of electrons in the orbitals constituting the neutral target molecule based on a Hartree-Fock description.
\end{enumerate}

\subsection{The scaling parameter $K_j$}
The value $K_j$ is an empirical scaling which accounts for the acceleration of the incident electron in the attractive potential of the target. As it relies on the Born approximation, without this scaling, the BEB model would significantly overestimate the ionisation cross section. Therefore, the $K_j$ scaling buttresses the success of the BEB model by effectively lowering the cross section at energies near threshold (where the Born approximation fails). 

At the same time, $K_j$ is the Achilles' heel of the BEB because there is no universal formula for estimating $K_j$ that would befit all sorts of impact ionisation cross sections. Instead, several semi-empirical variants have been proposed in order to extend the applicability of the BEB to a variety of targets: light/heavy atoms \citep{Guerra2012}, valence/core orbitals \citep[eqs.(6--7)]{Santos2003}, neutral/ionised targets \citep[§VI.]{Kim1994} or atoms/molecules \citep{Kim1997}. We will only mention the two following semi-empirical approximations:

\begin{description}
	\item[Burgess denominator :] proposed by \citet{Vriens1966} when taking into account that the orbital electron bound by $B_j$ has a kinetic energy $U_j\equiv \langle p_j^2/2m_\text{e} \rangle$ corresponding to its average squared kinetic momentum.
	\begin{equation}
		K_j = (U_j + B_j)/n_j \; . \label{eq:K-be}
	\end{equation}
	The kinetic energy term $U_j$ emerges as we average on the bound electron's orientation in the binary-encounter model. The division by the principal quantum number $n_j$ was added later in \citet[§III.B:p.1234]{Huo1999} as a correction for ionisation from outer shells of heavy nuclei ($n_j\geq 3$) [for atomic targets].
	\item[Screening correction :] proposed by \citet{Guerra2012} in order to replace a tentative modification of the BEB scaling for inner-shell ionisations by \citet[eq.(6)]{Santos2003}. Their new formula reads as:
	\begin{equation}
		K_j = 2\text{Ryd} \left(a \frac{Z_{\text{eff},n\ell j}}{2n^2} + b \frac{Z_{\text{eff},n'\ell' j}}{2n^{\prime2}}\right) \; . \label{eq:K-sc}
	\end{equation}
	The effective nucleus charge $Z_\text{eff}$ is obtained from the data of \citet{Clementi1963} as $Z_{\text{eff},1s} = 0.9834Z-0.1947\,, \, Z_{\text{eff},2s} = 0.7558 Z-1.1724$ and $Z_{\text{eff},n \ell j} = Z - \sum_{i=1}^{n-1}2(i-1)^2-2\ell^2$ otherwise. The full screening is a linear combination (set as $a=0.3$ and $b=0.7$) between the effective screenings of the current ($n\ell$) and the next occupied ($n'\ell'$) subshells (higher in energy). Note that for the valence subshell, the model sets: $n'\ell'=n\ell$. Additionally, for molecules, the principal quantum number $n$ is not a good number and may be used only on inner (non-valence) orbitals.
\end{description}

Both expressions try to estimate the kinetic energy $K_j$ of the incident electron at the place where ionisation takes place. Obviously, this classical view of ionisation has limitations; as the dominant dipole interaction is of long range and thus one may not ``locate'' the incident electron. Nevertheless, formally, one may say that the BEB model relies on the first (plane wave) Born approximation, which is known to always overestimate the interaction, since it does not consider distortion of the incident electron wave due to the potential of the target. The $K_j$ kinetic energy may thus be formally considered as an effective correction to the incident electron energy due to its distortion in the potential (including correlation with bound electrons) throughout the whole interaction. Further discussion on this scaling may be found in \citet[§B.1.a]{Tanaka2016} and references therein.

Since the purposes of the two scalings defined above are complementary, in section-\ref{sec:app} we will use eq.~(\ref{eq:K-be}) for valence shells and eq.~(\ref{eq:K-sc}) for core $1s$ shells (whether atomic or molecular) in the following study. In any case, the estimation of $K_j$, whether it stems from a binary-encounter approximation as in eq.~(\ref{eq:K-be}) or a screening estimation as in eq.~(\ref{eq:K-sc}), is independent from the intervention of the ionisation threshold $I_j$ in the bracketed term of eq.~(\ref{eq:beb}).

\subsection{The dipole oscillator strength} \label{sec:dos}

The assumption c) on the sum of the dipole oscillator strength density calls for a more detailed explanation. In the Bethe theory of inelastic collisions, the asymptotic expressions of both integral electronic excitation and ionisation cross sections are found to be proportional to a quantity named as the dipole oscillator strength and noted $f$ \cite[eqs.(4.18),(4.29)]{Inokuti1971}:

\begin{eqnarray}
	\sigma_{\mathrm{exc},j} \sim \frac{f_{\mathrm{exc},j}\mathrm{Ryd}}{W_j} \frac{\ln \varepsilon}{\varepsilon} + O(\frac{1}{\varepsilon})  \; , \\
	\sigma_{\mathrm{ion},j} \sim M^2_{\mathrm{ion},i} \frac{\ln \varepsilon}{\varepsilon} + O(\frac{1}{\varepsilon}) \; . \label{eq:ion-asymp}
\end{eqnarray} 

The labels $j,i$ are mere indexes for different discrete excitations and ionisation channels respectively. The quantity $W_j$ is the energy threshold for an electronic excitation indexed $j$. The electronic excitations for which $f_{\mathrm{exc},j} > 0$ are qualified as (optically or dipole) \emph{allowed}, whereas the rest are \emph{forbidden} and their asymptotic behaviour on the kinetic energy of the primary electron $\varepsilon$ falls faster ($\sim 1/\varepsilon$ or steeper).

The quantity $M^2_{\mathrm{ion},i}$ (known as the ``dipole-matrix-element squared'') is an integral over the kinetic energy of the secondary electron $w$ from the ionisation channel $i$, obtained with the \emph{differential} (or density) dipole oscillator strength \citep[eq.(19c)]{Tanaka2016}:

\begin{equation}
	M_i^2 = \int_{0}^{\infty} \frac{\mathrm{Ryd}¸}{w+I_i} \frac{\mathrm{d}f_{\mathrm{ion},i}}{\mathrm{d}w} \mathrm{d}w \; .
\end{equation} 

In addition to the Bethe approximation, the assumption b) above of the BEB devises a profile for the dipole oscillator strength density as \citep[eq.(46)]{Kim1994}:
\begin{equation}
	\frac{\mathrm{d}f_{\mathrm{ion},i}}{\mathrm{d}w} = \frac{N_i I_i}{(w+I_i)^2} \; , \label{eq:df/dw}
\end{equation}
where $N_i$ is the integral value of the oscillator density. Then, in the BEB, the value $M^2_{\mathrm{ion},i}$ equals $N_i\mathrm{Ryd}/2I_i$ \citep[eq.(49)]{Kim1994}.

The dipole oscillator strengths (and their integrals) are all linked and constrained by an equation known as the Thomas-Kuhn-Bethe sum rule and amounting to the total number $N$ of electrons bound by the target (atom or molecule) \cite[eq.(16)]{Tanaka2016}:

\begin{equation}
	\sum_{j} f_{\mathrm{exc},j} + \sum_i  \underbrace{\int_0^{\infty} \frac{\mathrm{d}f_{\mathrm{ion},i}}{\mathrm{d}w} \mathrm{d}w}_{\equiv N_i} = N \; . \label{eq:bethe-sum}
\end{equation}

Finally, the BEB's assumption c) on the Thomas-Kuhn-Bethe sum consists in setting $\sum_i N_i = N$ \citep[eq.(49) with $Q=1$]{Kim1994}, and henceforth $\sum_j f_{\mathrm{exc},j}=0$.  

Of course, this is not the case in practice and therefore this implies that the BEB should, in principle, asymptotically tend to overestimate the total ionisation cross section since it is encompassing all optically allowed excitations; whether they actually lead to ionisation of the target (like autoionisations) or not. In this way, we note that the BEB includes optically allowed autoionisations \emph{implicitly} (through the contribution of their dipole oscillator strength which affects the asymptotic region).

Notwithstanding, if we consult the various targets onto which the BEB has been applied, we do not observe any asymptotic overestimation. Indeed, the cyan curves on figure~\ref{fig:BEB} below reproducing the original total ionisation BEB cross sections align generally well with the experimental data. The examples matching best are N, CO, N$_2$, O$_2$, CO$_2$ and NO$_2$.

Asymptotically, the values at a few hundreds of eVs should not be affected by the scaling $K_j$ which, for the targets selected, stays $\lesssim$100 eV for ionisation from the valence shell accounting for most of the total cross section. Therefore, we surmise that this unexpectedly good agreement stems from the last assumption: namely the Hartree-Fock modelling of the target.

\subsection{The Hartree-Fock orbitals}

The total ionisation cross section of the BEB is constructed as a sum of ``orbital'' cross sections :
\begin{equation}
	\sigma_{\text{tot}}(\varepsilon) = \sum_{j=1}^{n_\text{orb}} \sigma_j(\varepsilon) \, , \label{eq:tot}
\end{equation}
Such denomination was proposed by the original authors, because these $\sigma_j$ (\ref{eq:beb}) are based on the theoretical orbital structure of atoms or molecules in the Hartree-Fock approach where each electron is bound to its orbital $j$ by a theoretical binding energy $B_j$. Additionally, only the lowest binding energy was replaced by its experimental value, so that the total cross section have a threshold consistent with experiments \citep[p.2966]{Hwang1996}. 

Such decomposition of $\sigma_\text{tot}$, however, is artificial and may not be used to estimate the sub-processes of ionisation from individual orbitals. This was well pointed out in \citeauthor{Hwang1996}'s conclusion \citep[p.2966]{Hwang1996}:
\begin{quote}
	[the BEB model] can be used -- for the total ionization cross section, not orbital cross sections --
\end{quote}

There are two reasons for this inadequacy:

\begin{enumerate}
	\item The Hartree-Fock model has its limitations and does not incorporate accurately the correlation forces between bound electrons. As a result, the Hartree-Fock values of the binding energies $B_j$ may differ significantly from experimental measurements $I_j$ \citep[table~1]{Duke1995}. This was used to point out some shortcomings of the Hartree-Fock approximation and molecular orbital theory for predicting the ordering of molecular orbitals binding energies \citep[e.g. for N$_2$:][]{Cade1966}. 
	\item The modelling of ionisation as a removal of an electron in an occupied orbital relies on the so-called generalised Koopman or ``frozen orbital'' approximation. The generalised \citet{Koopmans1934} approximation states that the binding energy $B$ of an electron in a Hartree-Fock atom/molecule described by a single Slater determinant is equal to the ionisation threshold $I$ of that atom/molecule for the removal of that electron. Usually, a single Slater determinant is not sufficient to properly describe an atomic or molecular state. But the greatest error induced by the Koopmans approximation is to picture the removal of an electron meanwhile the other electron orbitals remain ``frozen''; i.e. that they do not relax upon the disappearance of a bound electron \citep[§5:p.53]{ESCA2}.
\end{enumerate}

In brief, Hartree-Fock and Koopmans approximations do not accurately represent the correlation between bound electrons. After examination, we realised that the theoretically calculated orbital binding energies using these approximations are generally overestimating the experimental ionisation thresholds: $B_j\gtrsim I_j$. As a result, the total BEB cross sections when using the theoretical $B_j$ are roughly proportionally lower than they would be using the experimental $I_j$ (because the magnitude factor in eq.~\ref{eq:beb} $\propto1/I_j$). 
Furthermore, the Koopmans approximation misses out an important aspect of partial ionisation processes: that removal of an electron from a concrete orbital actually opens up to a variety of ionisation channels, each at a different ionisation threshold.

In the present paper, we propose a decomposition of the BEB (shown in table~\ref{tab:long}) into cross sections associated to specific ionisation channels with \emph{experimental} values of ionisation potentials $I$ as identified in photoelectron spectroscopy \citep{ESCA2}. In the next section, we introduce what such ionisation channels are.

\section{Ionisation channels}

From the discussion above, we know that having the BEB rely on the Koopmans approximation is restrictive because ionisation to an excited ionic state does \emph{not} merely correspond to the removal of an electron from an occupied orbital. To model accurately the underlying complexity of the ionisation process is a very difficult task. Ideally, one should aim to a complete description of all the processes that can be identified in the photoionisation spectrum -- and therefore participate in the dominant dipole term of the electron impact ionisation. Nonetheless, semi-empirical models, such as the BEB have proven to be simple yet fairly accurate representations of the \emph{total} ionisation cross section. We argue hereby that \emph{partial} ionisation to excited ionic states should be modelled in the BEB model with regard to the ionisation peaks or more generally the prominent features observed experimentally through photoionisation spectroscopy \citep{ESCA1}. In this way, the application of the BEB model does not need to restrict itself only to targets for which the lower cationic states are
well described by the removal of a single electron from the ground state configuration built from Hartree-Fock orbitals. 

Generally speaking, ionisation induces a perturbation on the orbital structure of the ion which we may exemplify in three types of situations:
\begin{enumerate}[a)]
	\item Coupling: for open-shell targets, ionisation from a closed subshell leads to different coupling possibilities of the orbital and spin moments with the electrons in the open shell. For instance, ionisation from the $2s$ orbital of the carbon atom can lead to a $^{4}\,{P}$, $^{2}\,{D}$, $^{2}\,{S}$ or $^{2}\,{P}$ ionic $\text{C}^+$ state. Another example is illustrated by the warmly (orange, yellow, brown) coloured lines of different stroke patterns representing the states ($\, ^{4}S^o$, $\,^{2}D^o$, $\,^{2}P^o$) of ionised atomic oxygen after ejection of a (2p) electron on the top graph of figure~\ref{fig:partial-BEB}.
	\item Reallocation: some of these states (the doublets) may only result from a subsequent reallocation of electrons in the orbitals of an open subshell ($2p$). The metastable singlet states of N$^+\,^{1}\,{D}$ and $^{1}\,{S}$ are other good examples, which require that one of the two remaining electrons in the $2p$ orbital (after ionisation) flip its spin projection. A change in electron spin is most probably due to exchange with the incoming electron \citep[p.11]{eMolColl} though (rather than a spin-flip interaction). 
	\item Shake-up: at high incident electron energies, it is possible that, in addition to ejecting an electron from its orbital, other electrons from the valence shell be excited to other orbitals. In X-ray photoelectron emission, this is known as ``shake-up'' \citep[p.26]{ESCA2}. An example of ``shaken-up'' ionic states are the B$\,^2\Sigma_u^+$ and C$\,^2\Sigma_u^+$ states of N$_2^+$ which are configuration mixtures between a removed $2\sigma_u$ electron and an excited electron in the $1\pi_g$ orbital \citep{Lorquet1972}.
\end{enumerate}

The connection between \emph{photo}electron spectroscopy and electron impact ionisation is established by the dipole oscillator strength in the Bethe theory. We thus expect that states responsible for prominent bumps in photoionisation spectra will also have important contributions in impact ionisation cross sections. Conversely, weak features will remain negligible. Below, we discuss how we can adapt the BEB model to take into account the aspects aforementioned. 

\begin{description}
 \item[Ionic state multiplicity.] In light atoms, the break of degeneracy in the energy levels from the Russell-Saunders $LS$ coupling is very weak. As a consequence, the relative contribution of ionisations to excited ionic states from the same valence subshell were proposed by \citet{Kim2002} to be weighted by the statistical multiplicity of the ionic states. For a state labelled as $^{2S+1}L$ (e.g. O$^+\; ^{4}\,{P}$ with $S=3/2$ and $L=1$) the multiplicity $m_p$ is given by $(2S+1)(2L+1)$. These statistical weights rely on combining initial states which can be treated as degenerate at incident electron energies much higher than the respective partial ionisation thresholds (away from resonances and couplings) \underline{and} only among ionic excitations which involve the same degree of interaction (dipole-allowed, spin-forbidden, etc.) \citep[p.12]{Price1981}. Thus, the parameter $N$ becomes a ``participation'' weight of electrons in the channel (rather than the number of electrons on an occupied orbital). For molecules, $N$ is weighted likewise according to the multiplicity of each final ionic state. One can consult the values of $N$ that we calculated for various ionisation channels in table~\ref{tab:long} at the end of this article.
 
 \item[Spin-forbidden ionisation channels.] When evaluating the participation weight $N$, one must also be cautious of whether the ionisation involves flipping of an electron's spin. Ionisation events accompanied by a change of electron spin in the occupied orbitals, are analogous to optically spin-forbidden excitations except that they are accompanied with an emission of an electron in the unbound continuum. They do not participate in the Thomas-Kuhn-Reich-Bethe sum rule as their differential oscillator strength distribution should be zero. This implies that such ionisation channels will not be visible on photoelectron spectroscopy and consequently, we can expect that the cross sections by such ionisation events be small compared to optically-allowed channels. 
 
 Accurate B-spline $R$-matrix calculations by \citet{Wang2014N} of spin-forbidden ionisations from N$\,{}^4S$ to N$^+\,^{1}\,{D}$ and $^{1}\,{S}$ show that they contribute to less than \qty{1}\,{\%} of the total ionisation. Given the fact that spin-forbidden excitations have an asymptotic decay at $\sim 1/\varepsilon^3$ \citep[p.341:left column]{Inokuti1971}, we expect spin-forbidden ionisations to have a similarly steep decay. In the BEB model, this minor contribution may only come from higher order terms (steeper than $O(1/\varepsilon)$ in equation \ref{eq:ion-asymp}) than that of the dipole oscillator strength.  Given their rarity of occurrence, we shall dismiss these contributions to total ionisation. We will comment more on this case in section \ref{sec:app-atom} when we assess the BEB for atoms. 
 
 \item[Vertical ionisation thresholds.] Moreover, if the target is a molecule, its bond length and molecular shape may significantly change after being ionised. A notable example is the nitrogen dioxide (NO$_2$) which, upon removal of its singlet (6a$_1$) electron, changes to a linear ${}^1\Sigma_g^+$ equilibrium geometry \citep[p.452]{Baltzer1998} as its isoelectronic CO$_2$ molecule. Thus, molecular ionisation induces also a vibrational transition for which we have to distinguish the adiabatic energy threshold from the vertical (peak) energy threshold as explained in \citet[p.55]{ESCA2}. In this paper, following \citeauthor{Kim1997}'s conclusion \citep{Kim1997}, we always use the \emph{vertical} ionisation threshold.
 
 \item[Autoionisation.] For resonances, the BEB model actually accounts for the oscillator strength due to dipole-allowed excitations leading to autoionisation, so that their contributions are tacitly included (see section~\ref{sec:dos}). This is because the BEB was obtained assuming that the Thomas-Kuhn-Bethe sum of the oscillator strengths is entirely due to ionisation\citep{Kim1994}. As for autoionisations from dipole-forbidden excited states (thus absent in the photoelectron spectroscopy), their decay, as the incident electron energy grows, is much steeper than from dipole-allowed states; so their overall contribution is restricted to a small resonant region.
 
 \item[Branching ratios.]
 The complexity of the ionisation process implies also that there is not only one way to produce specific ionic species. Apart from direct ionisation, ions can also be produced by autoionisation. Multiply excited ions can be formed by shake-off (interaction with more than one electron) or Auger emission subsequent to a core ionisation (which, in light atoms, is more common than X-ray deexcitation \citep{ESCA1}). However, the experimental cross section for production of multiply charged ions starts at a much lower threshold than the binding energy of a $1s$ electron and is also significantly larger than ionisation from the K-shell (fig.~\ref{fig:partial-K}). Thus, multiple ionisation mechanisms are mostly imputable to complex electron-ion interaction beyond the single binary-encounter model and subsequent autoionisation. 
 In molecules, double ionisation almost always leads to a ``Coulombic explosion'' whereby energetic atomic ions are ejected apart \citep[p.155--157]{Berkowitz1979}. Because of the multiple possible processes subsequent to ionisation to an excited state in molecules, one requires \emph{branching ratios} to estimate the production of an ion fragment from a particular ionisation channel. Such endeavour was already undertaken in other studies \citep{Huber2019} and will not be covered here.
\end{description}

In light of this correspondence analysis between the BEB model and ionisation processes, we proceed in the following section to the application of experimental thresholds determined from photo-ionisation spectra as input parameters to the BEB model and comment on the result.


%

\section{Application} \label{sec:app}

We now apply the separation of the BEB model into partial ionisation channels from the valence shell and from core orbitals to four light atoms: C, N, O, F; four diatomic molecules: CO, N$_2$, O$_2$, NO; and four triatomic molecules: CO$_2$, H$_2$O, NO$_2$ and O$_3$; all which are of interest in cold plasmas in atmospheric gases. The identified processes are gathered in table~\ref{tab:long} together with:

\begin{itemize}
	\item $I$ : the corresponding experimental ionisation potentials from photoelectron spectroscopy;
	\item $K$ : scaling factors calculated with:
	\begin{itemize}
		\item eq.~(\ref{eq:K-be}) for valence-shell orbitals using experimental ionisation thresholds for ``binding energies '' ($B=I$) and kinetic energies $U$ from \citet{Kim2002} (C,N,O atoms), \citet{Hwang1996} (diatomic), \citet{Kim1997} (triatomic) and from the orbital wavefunctions of \citet{Koga1999} (fluorine);
		\item eq.~(\ref{eq:K-sc}) for the core $1s$ shell;
	\end{itemize}
	\item $N$ : participation of electrons in each channel, as mentioned above, according to the multiplicity rule of the final ionic state $(2L+1)(2S+1)$ and the electron occupation in the corresponding orbital. Contributions from spin-forbidden reactions (displayed for comprehensiveness and noted with a barred arrow $\nrightarrow$) are assumed negligible.
\end{itemize} 

The last column gives the references from which the ionisation channels and their thresholds were extracted.

\begin{figure}
	\includegraphics[width=\textwidth]{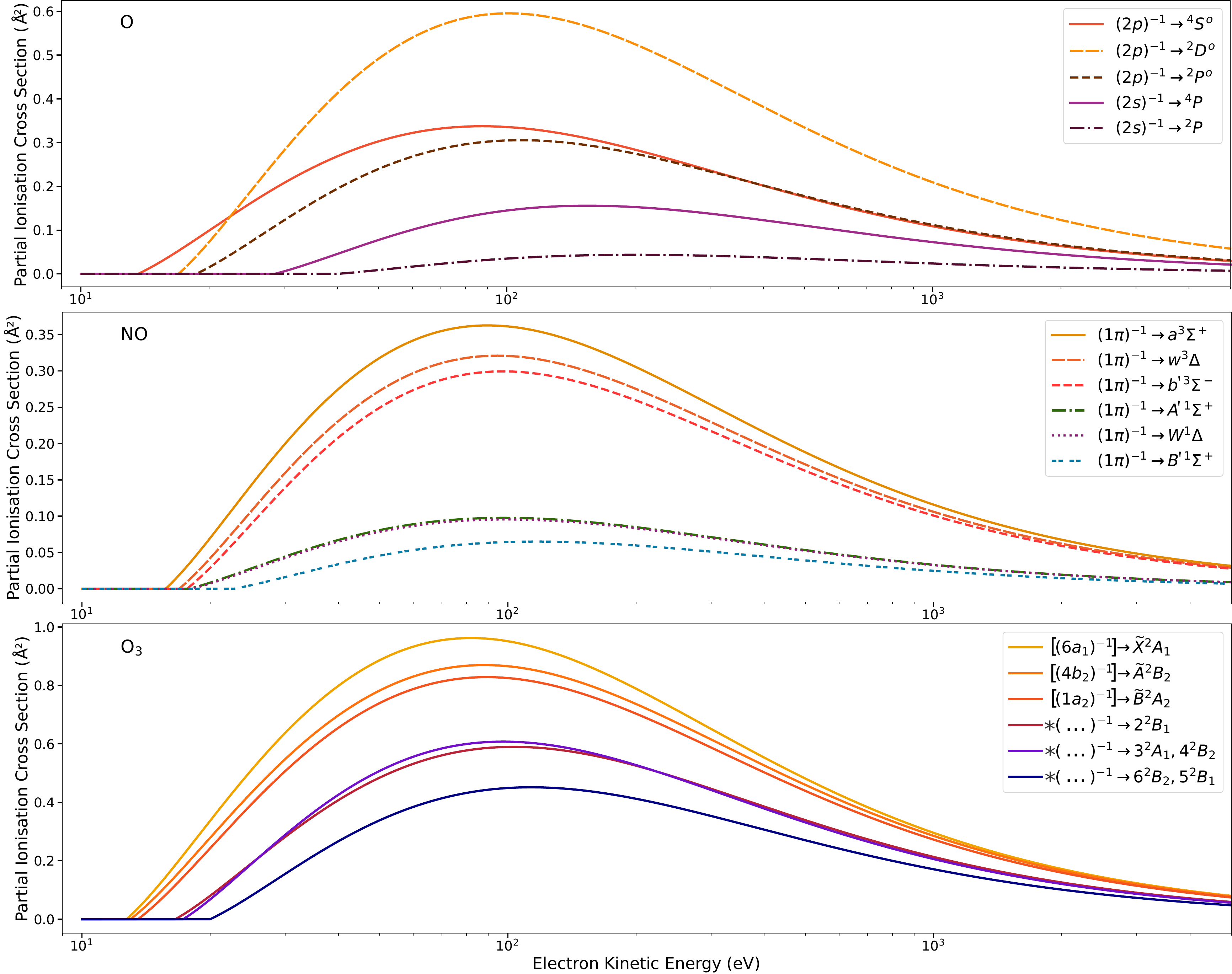}
	\caption{Partial ionisation cross sections obtained with the BEB model for atomic oxygen (top), nitric oxide (middle) and ozone (bottom) with parameters reported in table~\ref{tab:long}. Lighter colours are associated to outer orbitals (lower binding energies). Different line styles distinguish ionisations corresponding to a removal of an electron from a same orbital but with different ionic outcomes. Coupling between momenta of unpaired electrons leads to different ionic states whose multiplicities reflect their intensity as in atomic oxygen and the triplets and singlets in nitric oxide. The orbital picture does not apply well to ozone. Although the first three ionic states may still be identified with electron ejection from an orbital with a leading configuration [in square brackets], ionisations from deeper valence orbitals (\textcolor{darkgray}{*}) may only be identified with final ionic states whose many vibrational bands overlap over a certain region in the photoelectron spectrum, and which are bundled together to form the ``partial'' impact ionisation cross sections drawn on the bottom graph. \label{fig:partial-BEB} }
\end{figure} 

The BEB model with all identified partial ionisation channels is illustrated in figure~\ref{fig:partial-BEB}, for an atomic (O), diatomic (NO) and triatomic (O$_3$) gaseous target. Each line is associated to (a.)~the ejection of an electron from an orbital (in the Hartree-Fock picture) or to (b.)~an ionisation channel (in the more general picture). Different partial cross sections due to ionisation from the same orbital are represented with different line styles (solid, dashed, ...). Lighter colours are used for outer shells/smaller binding energies. We can observe the influence of the ionic multiplicity on the predicted partial ionisation cross section, and therefore a separation in magnitude between groups of different multiplicities as we see by the triplet and singlet ionic excitations of NO (middle graph). 

\begin{figure}
	\includegraphics[width=\textwidth]{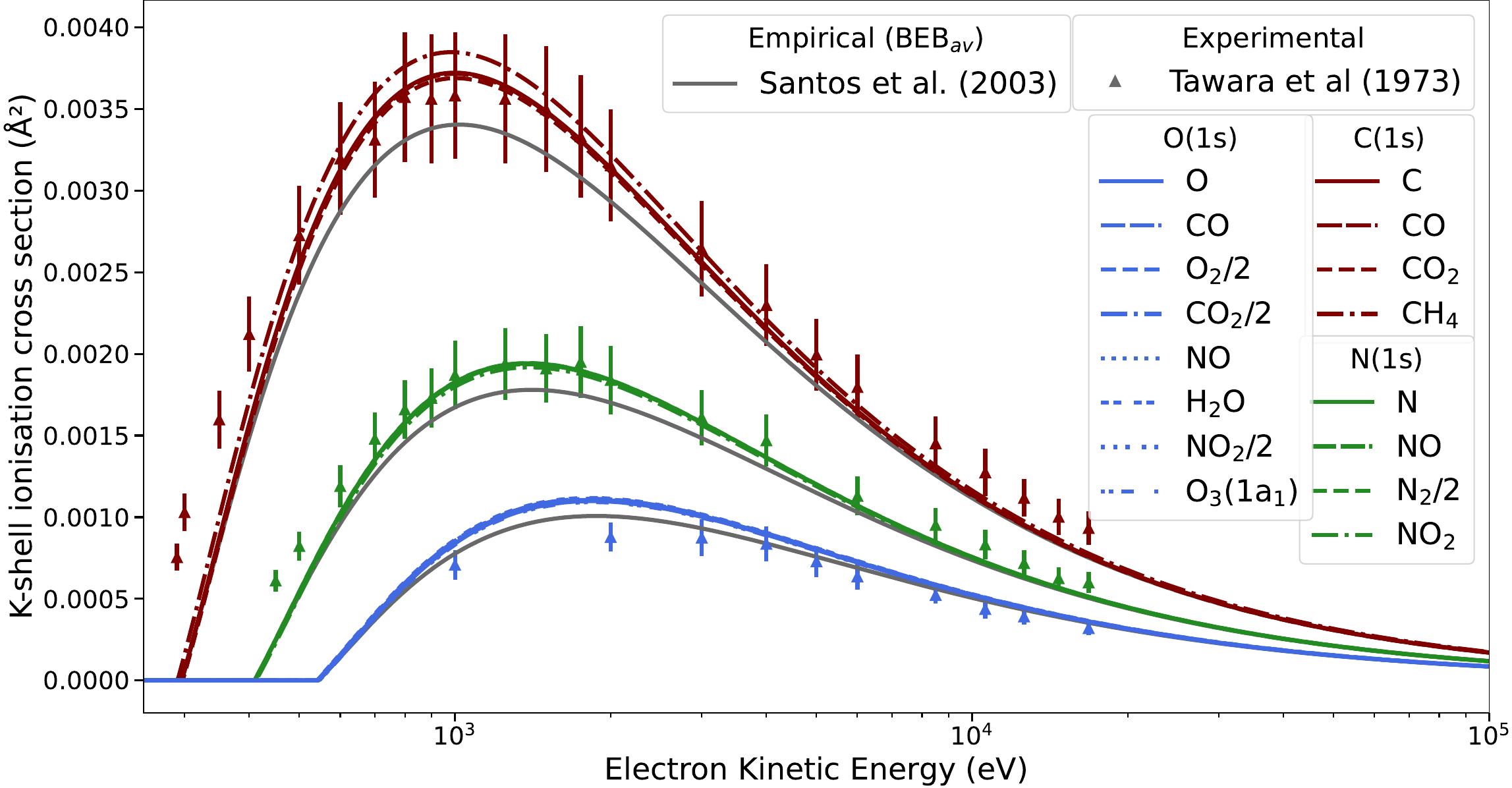}
	\caption{Partial ionisation cross sections from the 1s shell of C, N and O atoms in their free form and bound in various molecules. The 1s binding energies $I$ are reported in the first rows of table~\ref{tab:long} for each gaseous element. The scaling energy $K$ is calculated with eq.~(\ref{eq:K-sc}). Fluorescence factors used to convert experimental data are from \citet{Tawara1973}: $\omega_{C_{1s}}=\num{2.47e-3}$ \citep[table~I]{Honicke2023}, $\omega_{N_{1s}}=\num{4.35e-3}$, $\omega_{O_{1s}}=\num{6.91e-3}$ \citep{Liu2000} (see main text for explanation). \label{fig:partial-K}}
\end{figure} 

For better visibility, partial ionisations from the 1s (K) shell are displayed on a separate graph in figure~\ref{fig:partial-K}, where we stacked ionisations from C, N and O atoms according to the slightly varying 1s binding energies in different molecules. The greatest differences arise from 1s electrons bound to carbon in methane (CH$_4$) which was used in the experimental study of \citet{Tawara1973}. These experimental data were reported as X-ray production cross sections and require the fluorescence yield $\omega_{1s}$ in order to convert into impact ionisation cross sections. On this issue, note that the experimental K-shell cross sections displayed on \citet[fig.1]{Santos2003} for N are lower than in \citet[fig.11:p.46]{Llovet2014} using fluorescence yields reported in the EADL \citep{EADL}. For disambiguity, in the caption to our figure~\ref{fig:partial-K}, we specified the fluorescence yield $\omega_{1s}$ that we used. 

Aware of the inadequacy of the original BEB model for ionisation from core shells, \citet[eq.(6)]{Santos2003} proposed to patch the model with an empirical averaging of the scaling factor, the result of which is reproduced in grey solid curves. Overall, the $K$ estimated by screening eq.~(\ref{eq:K-sc}) gives a better agreement than the empirical averaging of \citet[fig.1--2]{Santos2003}. The slightly worse agreement for O is subject to debate on the value of the fluorescence factor used to convert the X-ray yield into cross sections. Additionally, we deem that the disagreement at energies near threshold is due to loss in accuracy of experimental results, since they \citep{Tawara1973} give non-zero values at energies below the ionisation threshold identified in photoelectron spectroscopy. Therefore, we advocate that the BEB model be used with $K$ as a parameter tabulated as we do so in table~\ref{tab:long}, rather than to rely on semi-empirical formulas based on other parameters (such as the kinetic energy $U$ of bound electrons). It would be challenging to find a universal analytical formula for $K$. On the other hand, we do not oppose proposals of calculating $K$ \textit{ab initio}, such as from the average ratio between cross sections from the plane wave and the distorted-wave approximations. 

\begin{figure}
	\includegraphics[width=\textwidth]{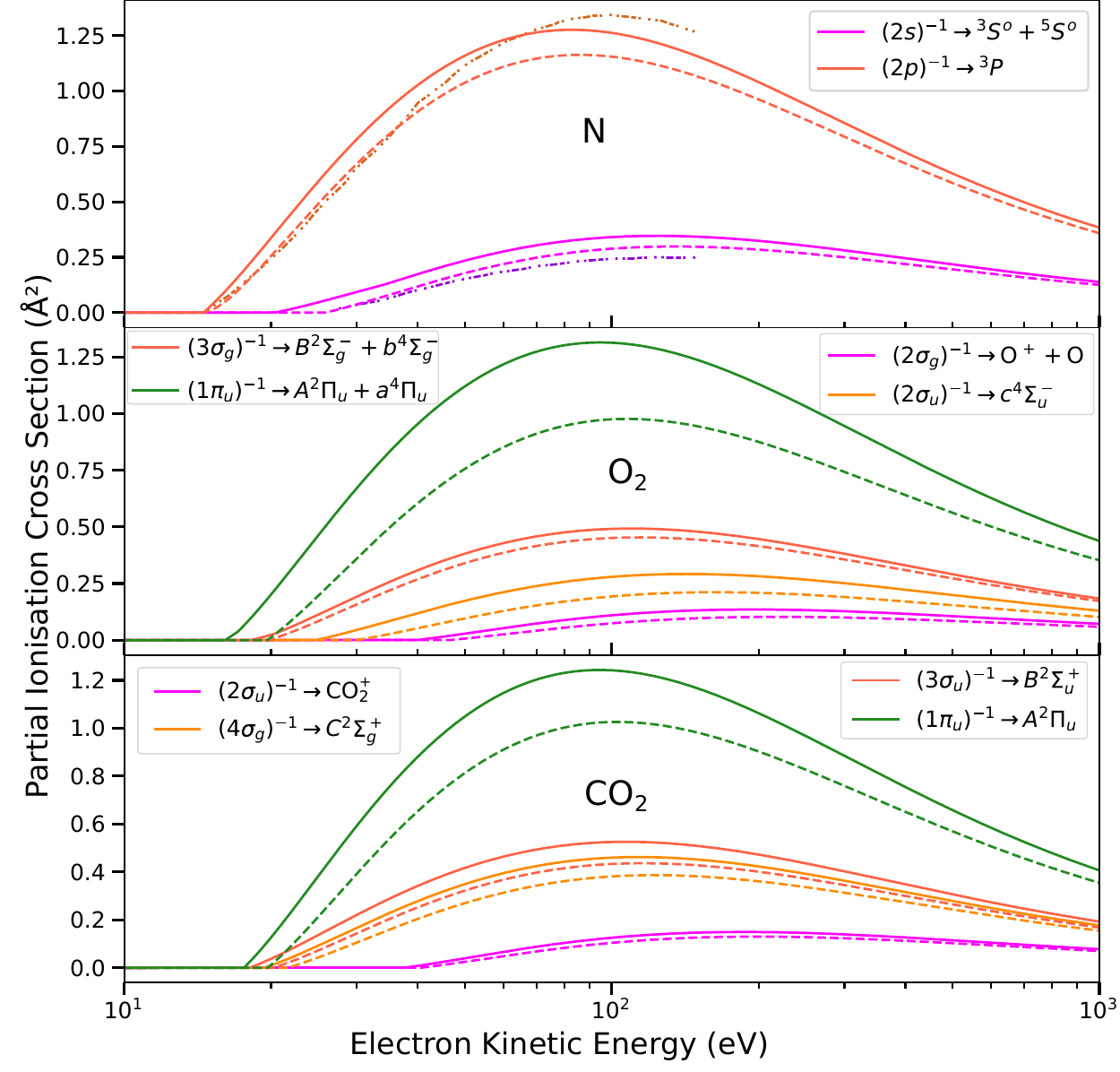}
	\caption{``Orbital'' ionisation cross sections obtained with the BEB model for atomic nitrogen (top), molecular oxygen (middle) and carbon dioxide (bottom). Solid lines (\textbf{---}) show the sum of partial cross sections from ionisation channels associated to removal from a molecular orbital with parameters reported in table~\ref{tab:long}. Dashed lines (\textbf{-\--\--}) of the same colour correspond to original BEB model with theoretical thresholds from Hartree-Fock calculations. The shattered lines (\textbf{$\cdots$--\---}) show the BSR calculations for N from \citet[figure~7]{Wang2014N}. Higher ionisation thresholds directly imply roughly proportionally lower cross sections as can be identified from eq.~(\ref{eq:beb}). This difference permains on the total ionisation cross section in fig.~\ref{fig:BEB}. \label{fig:BEB-comp} }
\end{figure}

When comparing with the original BEB model based on Hartree-Fock binding energies, one must recede to the removal from an orbital picture, and add all ionisation channels which originate from the same orbital. In figure~\ref{fig:BEB-comp}, one can see clearly that, for the three targets selected (N, O$_2$ and CO$_2$), all theoretical binding energies overestimate the experimental ones (the thresholds of dashed lines are all at higher energies than of solid lines of the same colour). As a direct consequence, the partial cross sections resulting from theoretical binding energies are scaled down compared to our newly calculated ones. 

\begin{figure}
	\includegraphics[width=\textwidth]{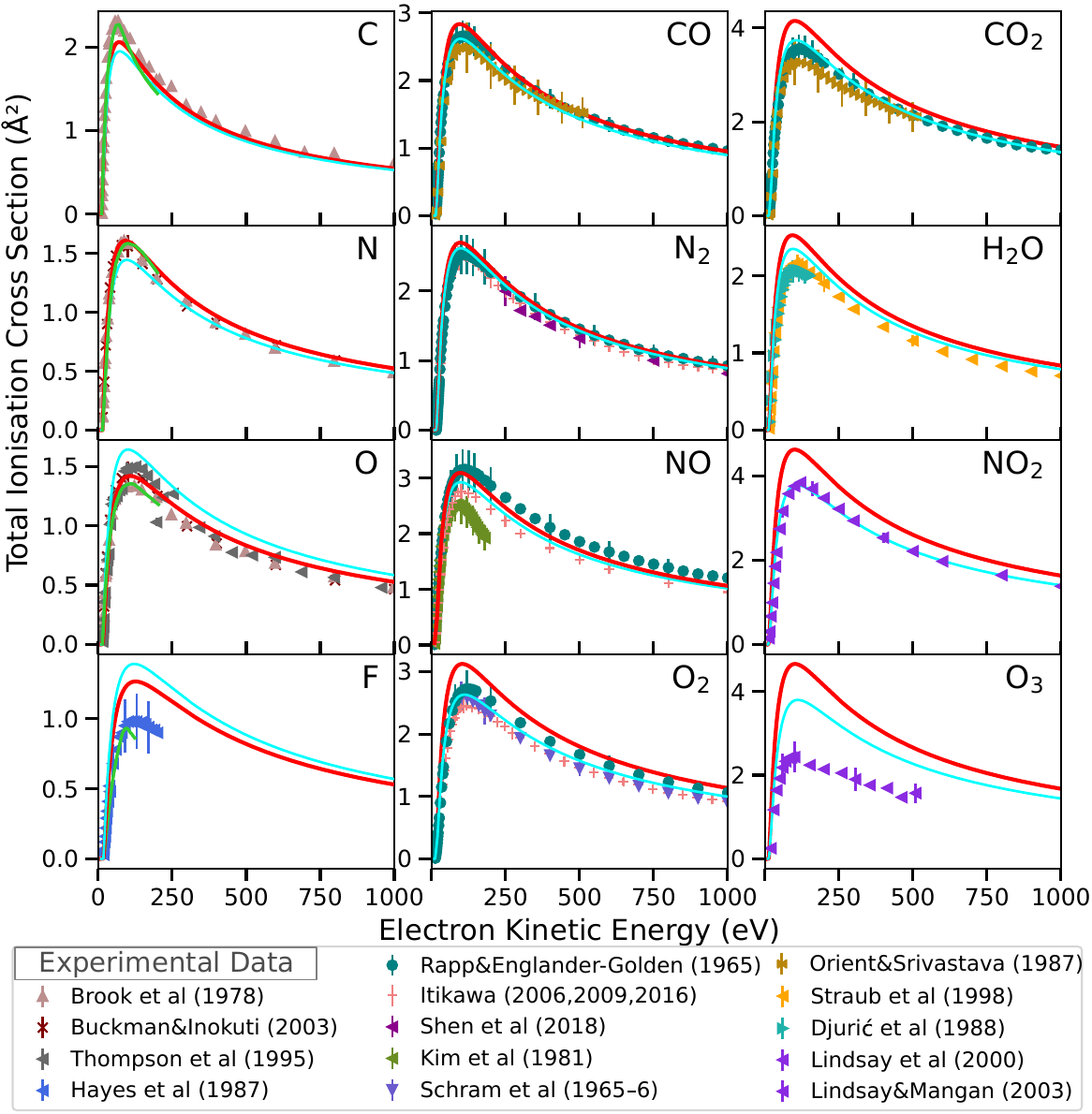}
	\caption{Sum of partial ionisation cross sections from the BEB model with experimental ionisation thresholds $I$ (table~\ref{tab:long}) in red solid (\textcolor{red}{\textbf{---}}) and original Hartree-Fock theoretical binding energies in cyan (\textcolor{cyan}{\textbf{---}})  compared with experimentally measured total ionisation cross sections for four atoms, four diatomic and four triatomic molecules. For atomic targets (left column), BSR calculations are shown in light green (\textcolor{green}{\textbf{---}}) from \citep{Wang2013C,Wang2014N,Tayal2016O,Gedeon2014F}. Experimental data are from \citep{Brook1978,phe-Atom-2,Thompson1995,Hayes1987,Gaudin1967,Schram1965,Schram1966,Almeida1995,Rejoub2002,Rapp1965,Orient1987,Itikawa2006,Itikawa2009,Itikawa2016,Shen2018,Kim1981,Djuric1988,Straub1998,Lindsay2000,Newson1995,phe-Molc-5.1}. The alignment of the source labels echoes the 4$\times$3 grid, although some sources give data for many different targets. \label{fig:BEB} }
\end{figure}

Finally, the total ionisation cross sections (in solid red), obtained as a sum of all the partial cross sections such as those presented in figure~\ref{fig:partial-BEB}, are compared to experimental total ionisation cross sections in figure~\ref{fig:BEB}. The success of the BEB model in reproducing the \emph{total} ionisation cross section was already well established in a series of publications \citep{Kim1994,Hwang1996,Kim1997,Kim2002}. Nevertheless, yon calculations (reproduced in solid cyan lines) relied on theoretical Hartree-Fock orbital binding energies which do not concord with the experimental thresholds for ionisation to excited states. The current figure~\ref{fig:BEB} shows that for some targets, good agreement can be preserved when using experimentally determined ionisation thresholds. For other targets, the BEB model ceases to give an accurate representation of the total ionisation cross section implying that a further scaling is necessary. 

Below, we comment on the total and partial cross sections of ionisation obtained for each of the targets considered. Based on information available in the literature, we explain how we identified each of the ionic states and whether these may be considered as pure states with holes or mixtures.  

\subsection{Atoms} \label{sec:app-atom}

Overall, the agreement of the sum of BEB partial cross sections with experimental data from atomic targets is good, except for fluorine. On left column of figure~\ref{fig:BEB} we also display the BSR calculations from \citet{Zatsarinny2013} in light green. It is worthwhile to mention that in the experimental measurements of \citet{Brook1978}, metastable excited atoms entered the collision chamber. However, the data used in our plots is the corrected cross sections after subtracting the extrapolated signal from these metastable states (whose ionisation threshold is lower than the ground state). Therefore, we cannot exclude that some discrepancies might be also due to inaccuracy of the experimentally derived cross sections.

\paragraph{C.} When it was first modelled with the BEB, \citet[fig.~1]{Kim2002} argued that the shortcomings of the BEB ionisation cross section could be compensated by including contributions from autonisations. While we do believe that this assumption might be correct, we argue that, before adding autoionisations, one ought to first subtract the part of the oscillator strength imputable to autoionisations from the BEB model. Indeed, the asymptotic form of the separately calculated autoionisation (ai) cross sections follows $\sigma_{\mathrm{ai}} \sim f_{\mathrm{ai}} \ln \varepsilon/ \varepsilon$. Then, it follows from the discussion in section \ref{sec:dos}, that the dipole part of the BEB should be scaled as $(N-f_{\mathrm{ai}})/N$. 

In practice, this means to replace the BEB model by the so-called ``BE$Q$'' model where the $Q$ parameter plays the role of the (normalised) Bethe sum. This would lower the corresponding BE$Q$ cross section, but since our newly calculated cross section (red solid line on fig.~\ref{fig:BEB}) is larger than the previously used BEB, the scaling down by $Q$ could then make the two (solid red and cyan) curves match and preserve the agreement seen on \citet[fig.~1]{Kim2002}. Nonetheless, \citet[p.6~left]{Wang2013C} argue that this agreement is fortuitous due to an overestimation of the oscillator strengths used for the autoionising states. This should be verified in a more detailed study incorporating the dipole oscillator strengths which is beyond the scope of the present study (limited to the pure BEB model). 

For partial ionisations from the K-shell (fig.~\ref{fig:partial-K}), we used the experimental data of the 26$^\text{th}$ line reported in table IV of \citet{Bruch1985} and added the ionisation potentials (C$\,$I$\approx$\qty{11.26}{eV} and C$\,$II$\approx$\qty{24.383}{eV}) reported in the NIST database \citep{NIST-ADS}. \citet{Santos2003,Guerra2012} extracted their data from the same source, but we lack information about how the numerical value of the binding energy was derived.

\paragraph{N.} The situation is similar to carbon; updating the thresholds raises the total BEB cross section. This casts doubt on the previous conclusion that \citeauthor{Brook1978}'s \emph{adjusted} experimental data \citep[table~5]{Brook1978} is notwithstandingly plagued with 30\% of N$\,^{2}D$ as suggested in \citet[fig.~3]{Kim2002}. We have two strong supporting reasons to claim that this was not the case: 
\begin{enumerate}[i.]
	\item The experimental data of \citet[table~5]{Brook1978} is \emph{already} accounting for the spurious signal from metastable species. From their measured cross section (``$Q_m$''), they subtracted ionisation from metastable states using an ``apparent'' cross section $Q_a^*$ extrapolated with a simple power-law ``constant/$\epsilon$'', and obtained thus a ``true'' cross section $Q$ (reported in the middle column of their tables). While it is not sure how crude that estimation is, when comparing with experimental data, \citet{Kim2002} ought to have used the measured cross section $Q_m$ and not the corrected cross section to validate their argument about the mixture of metastables (from our close examination of their figure, it seems that they have used the adjusted value $Q(E)$ in \citep[table~5]{Brook1978}).
	\item The original cross section of \citet{Kim2002} (cyan fig.~\ref{fig:BEB}) is low because they statistically weighted total ionisation to two spin-forbidden ionisation channels namely: $(2p)^{-1}\nrightarrow \text{N}^+\,^{1}\,{S}$ (18.59 eV) and $(2p)^{-1}\nrightarrow \text{N}^+\,{}^{1}\,{D}$ (16.43 eV) higher than the ionisation from ground channel at 14.53 eV. Later, B-spline calculations from \citet{Wang2014N} found out that the contribution from the aforementioned spin-forbidden ionisation channels is negligible (> 1\%). As a result, they obtained a larger ionisation cross section from the 2p orbital than we did.
\end{enumerate}
A comparison of ionisation channels from 2p and 2s orbitals of BEB is given on the top graph of figure~\ref{fig:BEB-comp} where we show additionally in shattered linestyle the BSR calculations from \citet[fig.~7]{Wang2014N}. 
\begin{description}
	\item[(2p)$^{-1}$:] The BSR results are closer to the BEB with original thresholds (with statistical weighing to forbidden-states included) at energies up to 50 eV, but then the BSR calculations exceed the BEB model even with updated thresholds. Assuming that the BSR calculations are accurate, that would mean that the BEB would need to have its scaling ($K$) revised as in eq.~(\ref{eq:K-be} since it is effective only at energies near threshold. 
	\item[(2s)$^{-1}$:] The BSR does not have the same threshold as the one we found for N$^+ (2s2p^3)\, {}^{3}S^o$ at 20.34 eV from the NIST database \citep{NIST-ADS}. This results in a higher BEB cross section compared to the BSR. Further clarification would be needed in order to understand why the thresholds do not concord. 
\end{description}
In the meantime, we extend \citeauthor{Wang2014N}'s critique of the BEB results, by stating that the model needs revision on the thresholds used and the ionisation channels open.

\paragraph{O.} For oxygen, we are unable to reproduce the original BEB from \citet{Kim2002} because they did not disclose the thresholds used to generate the two metastable ions O$^+\,^2{D}^o$ and O$^+\,^2{P}^o$. When we use our thresholds, we obtain the red solid curve on figure~\ref{fig:BEB} which already overestimates the cross section in the asymptotic region and does not underestimate it below 100 eV to the extent observed on \citep[fig.~4]{Kim2002}. Once again, we do not deny that atomic oxygen has several autoionisation channels, but if these are to be included separately, the BEB must be downscaled according to the part of the dipole oscillator strength taken by these autoionisations (same argument as for carbon). The BSR calculation digitised from \citet{Tayal2016O} reproduces well the measurements of \citet{Brook1978}. The authors argued that differences of the BEB with BSR is due to the modelling of autoionisating excitations. 

\paragraph{F.} For fluorine, the overestimation of the cross section results from a too large optical oscillator strength of the ionisation channels. Overall 12\% of the Bethe sum is taken by discrete excitations \citep[table~II]{Gedeon2014F}. Then, better results can be obtained when a more accurate model of the dipole oscillator strength is used \citep[§3.2,fig.~9]{Brook1978,Kim1994}. Although it seems that fluorine lies beyond the applicability of the BEB model, BSR calculations from \citet{Gedeon2014F} agree well with the experimental data of \citet{Hayes1987}.


\subsection{Diatomic molecules}

Except for O$_2$, the difference on the total ionisation cross section between the original and revised BEB is minor, while the thresholds can differ by 10\%. 
For each molecule below, we discuss the various ionisation channels identified in relationship with the orbital structure and the likelihood of dissociating after ionisation. Where possible, we provide information on branching ratios of dissociation and predissociation of molecular ions obtained from clastic photoionisation cross sections. In general, the yield of doubly ionised whole molecules (which depends also on their lifetimes), though non-zero, is negligible compared to dissociation into ionic fragments \citep[p.5-20,63]{phe-Molc-5.1}.

\paragraph{N$_2$.} The two closed-shell and isoelectronic molecules are N$_2$ and CO. They are the simplest to describe in terms of removal of an electron from a molecular orbital, although the B and C states of N$_2$ are not pure states but different configuration mixtures with an excited $2p\pi$ electron \citep[table~1]{Lorquet1972}. The C state predissociates from the vibrational level $v'\geq3$ \citep{Erman1976}. Removal of a $2\sigma_g$ electron most likely dissociates into N+N$^+$, whereas a hole in the core orbital provokes an Auger emission followed by a separation into N$^+$+N$^+$ \citep[p.216]{Berkowitz1979}. A comparison with the partial ionisation cross section for N$^+$ fragments of \citet{Straub1996} would reveal that there is a significant portion of N$^+$ currently unaccounted by theoretical predictions from ionisation channels. These could come from a post-collision interaction of a secondary electron emitted from the $2\sigma_u$ orbital, whose energy is high enough to excite the nitrogen ion beyond its dissociation limit at \qty{8.7}{eV} as suggested by photoelectron spectroscopy \citep[p.64]{ESCA2}.

\paragraph{CO.} Presently, carbon monoxide is the only diatomic molecule for which we maintained a one-to-one correspondence between ionisation channels and ejection from molecular orbitals (see table~\ref{tab:long}). Thus, the lifting of the red from the cyan curve on top of figure~\ref{fig:BEB} comes from a reevalution of the higher ionisation thresholds whose former overestimation rated roughly 10\%. 

Nonetheless, above the B$^{2}\,{\Sigma}^+$ state (ejection of 4$\sigma$ electron), the excited ionic structure becomes more complex and there are many weakly bound states which predissociate into C$^+$+O fragments \citep[p.229]{Berkowitz1979}. These cannot be simply modelled in the bound-orbital structure as presented here. The details about dissociation paths ensuing the removal of a $3\sigma$ electron is subject to discussion. From ionisation of core orbitals, the yield in atomic carbon and oxygen ions is assumed to be roughly equal. 

\paragraph{NO.} Due to having a single $2\pi$ electron, nitric oxide has the most diverse array of excited ionic states. Six states (three singlets and three triplets) may be formed by removing one $1\pi$ electron due to the different coupling possibilities with the lone $2\pi$ electron \citep[table I]{Huber1968}. This separation into singlet and triplets can be observed as two groups of orange curves on the partial cross sections represented on figure~\ref{fig:partial-BEB}--middle graph. The labelling of the highest lying B'$^{1}\Sigma^+$ singlet requires confirmation \citep[p.482]{CDM}. 

NO is a good example to show that, when the Hartree-Fock calculated binding energy (18.49 eV \citep[table~1]{Hwang1996})  of an orbital (1$\pi$) lies somewhere in the middle of a more complex subdivision of ionisation channels ($\in[15.73;22.73]$ eV), the difference in the total ionisation cross section between original (cyan) and revised (red) BEB remains smaller than for CO as observed on figure~\ref{fig:BEB}; despite the fact that there is a qualitatively greater difference between the original and revised modellings of NO than for CO.

The energy range of the different states with a hole in the $1\pi$ orbital overlaps with the states with a hole in the $5\sigma$ orbital. Therefrom, one can find a different ordering of these orbitals (5$\sigma$ and 1$\pi$) according to their ``binding'' energy when the electronic configuration of NO is presented in the literature (e.g. compare \citet[p.231,table I]{Berkowitz1979,Hwang1996} versus \citet{Huber1968}). Since, the vibrational bands of some of the $(1\pi)^{-1}$ and $(5\sigma)^{-1}$ states overlap, they can be  pragmatically grouped together under a common binding energy, but with different relative contributions \citep{Edqvist1971}.

Next, the two higher ionic states $c$ and $B$ coming from removal of a $4\sigma$ electron are actually suspected to be mixtures with a $(2\pi)^2$ configuration \citep{LefebvreBrion1971}. In any case, they predissociate predominantly into N$^+$ + O \citep[p.236]{Berkowitz1979}. Also, there are indications that removing a $3\sigma$ electron also favours a N$^+$ + O dissociation \citep[p.237]{Berkowitz1979}, though we do not know whether the states are repulsive or weakly bound and predissociate. 

\paragraph{O$_2$.} Molecular oxygen has a partially occupied $1\pi_g$ orbital and, as a result, ionisation from inner subshells can either produce quartet or doublets \citep[p.219]{Berkowitz1979}. The improper ordering of the $1\pi_u$ and $3\sigma_g$ orbital energies of the Hartree-Fock model \citep[table~1]{Hwang1996} is responsible for the significant overestimation of the corresponding ionisation threshold (green solid and dashed lines on the middle graph of figure~\ref{fig:BEB-comp}) and explains the major difference in the total ionisation cross section between original and revised BEB. 

Predissociation of oxygen starts from the $v'\geq4$ of b$^{4}\,{\Sigma}_g^+$, but is weak. It becomes stronger for the B state and the c$^{4}\Sigma_u^-$ can be considered to predissociate completely\citep[p.222]{Berkowitz1979}. Whereas ionisation from the core orbitals leads to Auger emission and dissociation, removal of a $2\sigma_g$ electron can both lead to formation of a very high-lying (bound) doublet state or a dissociating (repulsive) quartet state. There exist a series of excited ionic states which stem from more complicated configurations.

\subsection{Triatomic molecules}

Overall, the overestimation of the total ionisation cross section for triatomic molecules is worse than for diatomic ones (where the revised BEB model still gives very reasonable estimation). 

\paragraph{CO$_2$.} Carbon dioxide is the only linear triatomic molecule, the others (in their ground state) belong to the $C_{2v}$ symmetry point group \cite[see][§4.3.2]{AMO2}. A good summary of carbon dioxide's properties and electronic structure is available in \citet[§1.2]{Aresta2016}. Similarly as for CO, the difference of the original and revised BEB is due to a former overestimation over 10\% of the higher ionisation thresholds which can be seen on the bottom graph of figure~\ref{fig:BEB-comp}.

Of the first ionic levels, only the $\tilde{\text{C}}$ state seems to predissociate. 
The inner $3\sigma_g$ and $2\sigma_u$ orbitals lie very close in energy and are mixtures of O(2s)+C(2s) and O(2s)+C(2p$_z$) respectively \citep[p.146:table~5]{Allan1972}. Ionisation from these inner orbitals seem to lead predominantly to a C$^+$+ 2O dissociation \citep[p.255]{Berkowitz1979}. Removal of a 4$\sigma_g$ leads to the $\tilde{C}$ state which predominantly predissociates into O$^+({}^{4}\text{S})$ + CO$({}^1\Sigma)$, while higher vibrational states may predissociate into CO$^+$(${}^2\Sigma^+$)+O(${}^3\text{P}$) to a ratio of CO$^+$:O$^+$ about 1:3 \citep[p.253--4]{Berkowitz1979}. 

\paragraph{H$_2$O.} Water is the simplest molecule presented here having the $C_{2v}$ symmetry. The nomenclature of molecular states and orbitals applying to this symmetry can be revised in \citet[§4.3.3:p.257--9:table 4.2--3]{AMO2}. Because of its complex vibrational spectrum, ionisation of water from each orbital comprises a rich vibrational band of closely spaced narrow peaks \citep[p.4746:fig.1]{Karlsson1975}. Dissociation of the H$_2$O$^+$ ion starts after removal of a (1b$_2$) O--H bonding electron. The clastic cross sections relative to H$_2$O$^+$ are about $\lesssim$0.5 for OH$^+$:H$_2$O$^+$ and $\sim$0.12 for H$^+$:H$_2$O$^+$ \citep[p.244]{Berkowitz1979}. Removal of a (2a$_1$) electron causes complete dissociation with a H$^+$:O$^+$ ratio of 3:1 \citep{Tan1978}. Finally, ejection of a (1a$_1$) electron entails a rapid Auger decay and a subsequent Coulombic explosion leading plausibly to 2H$^+$ + O$^+$ \citep[p.246]{Berkowitz1979}.

\paragraph{NO$_2$.} The orbital symmetries associated to ionisation channels of NO$_2$ in table~\ref{tab:long} are following \citet[p.451]{Baltzer1998}. There are differences with the notation in \citep[tableĨI]{Kim1997} and also in the ordering of energy levels. The notable difference in the total ionisation cross section between original and revised BEB in figure~\ref{fig:BEB} is mainly due to the ionisation channels associated with 5a$_1$, 1b$_1$ and 3b$_2$ orbitals whose Hartree-Fock energies are actually about 18\% above the experimental ones, supplemented with a 10\% overestimation of other thresholds. Despite this inadequacy of binding energies, the agreement of the original BEB with the total ionisation cross section is mesmerising. We believe this agreement to be coincidental.

Like NO, nitric dioxide has a lone electron in its outermost valence shell (6a$_1$). Removing this lone electron makes NO$_2^+\,\text{X}$ linear and isoelectronic with CO$_2$. This consequential change of geometry spurs a very dense and rich vibrational spectrum whose adiabatic ionisation threshold may be as low as \qty{9.6}{eV} \citep{Clemmer1992}. Nevertheless, the onset of ionisation may be considered to start at the vibrational peak centroid around \qty{11.2}{eV} \citep[fig.~5]{Baltzer1998}. Ionisation from all orbitals below the highest occupied molecular orbital can generate either triplet or singlet states. Nonetheless, this splitting from holes in orbitals deeper than 4b$_2$ and 1a$_2$ is not visible on photoelectron spectra \citep[p.454:fig.~1]{Baltzer1998} and thus we group the split states together. Although most principal higher ionisation states still comply with the one-electron-hole picture, multiconfiguration calculations indicate that some states, such as (1b$_1$)$^{-1}\,\text{d}\,{}^3\text{B}_1$ result from significant configuration interaction \citep[p.584:table~1]{Chang2009}. The binding energies in inner valence orbitals are taken as broad peak positions in the photoelectron spectrum of \citet[p.455:table~2]{Baltzer1998}. 

The low dissociation threshold ($\approx$\qty{2.78}{eV}) of NO$_2$ causes all excited ionised states to have high dissociation branching ratios mainly fragmenting into NO$^+$ + O \citep[p.687:§IV]{Shibuya1997}. The relative production of NO$^+$ peaks at about \qty{20}{eV} to \qty{75}{\percent} \citep[fig.~5]{Au1997} and slowly seems to stabilises roughly at \qty{50}{\percent} at higher energies \citep[p.37:fig.~5]{Masuoka2004}.  However, state-specific branching ratios are found to vary significantly \citep[p.146--7]{Eland1998}. The main production of O$^+$ ions seems to come from the $(3\text{b}_2)^{-1}$ state \citep[p.146:§3.7]{Eland1998}. Here, we have to note that, although predissociation branching ratios may be equally applied to ionic species formed either by photoionisation or electron impact, it is not guaranteed that this analogy of clastic ratios applies equally well for ionisation channels leading to direct dissociation.

\paragraph{O$_3$.} Ozone is the most complex molecule presently introduced and the only one whose ground state may not be accurately described by a single Hartree-Fock configuration. It involves mixing of the $[...](1\text{a}_2)^{2}(4\text{b}_2)^{2}(6\text{a}_1)^{2}(2\text{b}_1)^{0}$ and $[...](1\text{a}_2)^{0}(4\text{b}_2)^{2}(6\text{a}_1)^{2}(2\text{b}_1)^{2}$ configurations \citep[§1]{Wiesner2003}. As a result, all the ionisation processes reported in table~\ref{tab:long} for O$_3$ (except for the core 1s orbitals) may not be interpreted as a simple hole in an orbital but result from more complex configuration interactions (thus the dashed arrows). Nevertheless, for consistency, we wrote the leading configuration as reported by \citep[table~3]{Wiesner2003} for the three first ionic states. The production of O$_2^+$ has an onset at \qty{13.2}{eV}, which coincides with the $(1\text{a}_2)^{-1}\,\tilde{B}\,{}^2\text{A}_2$ state \citep[fig.~2:inset]{Mocellin2001} who is thought to predissociate. All higher ionic states dissociate predominantly into O$_2^+ + $O. States above the $\tilde{B}\,{}^2\text{A}_2$, involving excitations from the deeper 5a$_1$, 1b$_1$ and 3b$_2$ orbitals, have non-leading mixed configurations and their vibrational bands overlap so that their contributions can be lumped together. In table~\ref{tab:long}, we selected the vertical energies of the bands (5), (6) and (7,8,9 grouped together), as reported by \citet[table~I]{Couto2006}. However, the assignment of the ionic symmetries is poorly documented and the references attributed do not actually discuss these bands. Thus, we put the labelling into squared brackets. 
The region above \qty{26}{eV} in the photoionisation of ozone is even more obscure, as we could not find any information involving the (3a$_1$) (2b$_2$) and (4a$_1$) inner orbitals. For completion, we took the theoretical values reported in \citet[table~II]{Kim1997}. Despite the many uncertainties about modelling partial electron impact ionisation from ozone, the particularly poor agreement of total electron impact ionisation cross sections seen on the last graph of figure~\ref{fig:BEB} should partly be due to inaccuracy in the experimental data which indeed give an unusually low cross section (compared with other triatomic molecules). This was attributed by \citet[§III.I]{Kim1997} to an inaccurate normalisation of experimental data of \citet{Newson1995}. Hitherto, as far as we know, this discrepancy has not been resolved.

Ozone challenges the BEB model based on an orbital decomposition of molecules as illustrated on the dotted labels on the bottom graph of figure~\ref{fig:partial-BEB}. Since the electron-hole picture collapses in the higher excited O$_3^+$ states, the attribution of the parameter $N$ becomes also uncertain. This exemplifies the need of a more generalised BEB model which is not based on the occupation number $N$ of orbitals (and ionic state multiplicity) but rather on the intensity of the (differential) dipole oscillator strength integrated over each band region. The ionisation cross section may still be decomposed into several partial ones with different thresholds, but the parameters $K$, $N$ and the labelling of the cross sections must be reconsidered.

\section{Conclusion and outlook}

We assessed the applicability of the binary-encounter Bethe model to the evaluation of partial ionisation cross sections from electron impact. We determined that its main limitation comes from its construction as a sum of ``orbital'' cross sections reflecting the structure of a single-configuration Hartree-Fock model, which overall gives a poor description of many of the molecules considered. This limitation is manifest when comparing the theoretical partial ionisation thresholds, systematically overestimating the experimental thresholds observed in photoelectron spectroscopy. 

We argue that, in order to yield partial cross sections of ionisation channels in equation (\ref{eq:tot}) consistently, all ionisation thresholds should be taken from experimental data using photo-electron spectroscopy. Unfortunately, this forfeits degradation of the overall performance of the BEB for representing total ionisation cross sections of molecular targets. Indeed for all molecules, the original BEB model using theoretically calculated binding energies from the Hartree-Fock model agrees better with experimental data than when using experimentally determined ionisation thresholds. 

Notwithstanding, such observation serves to elucidate how fortuitous compensation of errors endorsed the success of the simple BEB model. This was remarked already in the series of the B-spline R-matrix studies \citep{Wang2013C,Wang2014N,Tayal2016O,Gedeon2014F} for atomic targets. This observation changes previous conclusions about the performance of the BEB model. We wish to clarify the following:

\begin{enumerate}[I.]
	\item For atomic targets, spin-forbidden ionisations (see barred arrow $\nrightarrow$ on table~\ref{tab:long}) have negligible contribution. This affects the premise about the proportion of metastable states in the experimental data of atomic nitrogen.
	\item Contributions to the total ionisation cross section from all autoionising states are already encompassed by the BEB model because it assumes that all the Bethe sum (of the dipole oscillator strength) is entirely taken by ionisation processes. If one wishes to model autoionisations separately, one needs to consistently subtract the part of the dipole oscillator strength taken by the latter from the BEB model (which ceases to be BEB and turns into the BE$Q$ model with $Q<1$). This misunderstanding affected results on atomic carbon and oxygen. 
	\item A higher ionisation threshold induces an inversely proportional reduction of the partial cross section, principally affecting the asymptotic tail of the BEB. This explains why many of the molecules (O$_2$, NO$_2$, CO$_2$) which are known to have important optically allowed excitations (not leading to ionisation) could be modelled surprisingly well by the BEB even in the asymptotic region. Normally, one would expect to have an overestimation proportional to the part of the Bethe sum which had not be subtracted, as we now currently have. 
\end{enumerate}

The overall tendency of the binary-encounter-Bethe model to overestimate total ionisation is actually not unknown and can be corrected with a reduction of the optical oscillator strength encompassed in a parameter $Q$ used in the BE$Q$ model of \citet{Kim1994}. In the author's thesis \citep[§11.5.4]{Schmalzried2023}, we could accurately fit $Q$ to reproduce the total ionisation cross sections of atmospheric gases. Yet, we had not used the partial ionisation thresholds as identified here. In this paper, we wish warn readers against fortuitous successes of simple semi-empirical models. Only after using the experimental $I$ will fitting $Q$ open a promising avenue for representing partial cross sections with the binary-encounter-Bethe model. 
With these issues in mind, we propose that:

\begin{enumerate}
	\item The theoretical binding orbital energies $B$ be replaced by vertical ionisation thresholds $I$ identified in photoelectron spectroscopy as the energy at the maximum of sharp peaks or broad band centroids.
	\item The overestimation of the BEB model in the asymptotic region of the total ionisation cross section can now serve to identify the scaling due to the Bethe sum (of the optical oscillator strength) accounted by the more accurate BE$Q$ model.
	\item The value of the Bethe scaling parameter $K$ should depend on the process considered and may, but need not, be related to the orbital binding $B$ and kinetic $U$ energies as used in the Burgess denominator given in eq.~\ref{eq:K-be}. Indeed, \citet[p.2966]{Hwang1996} encouraged users to treat $U$ (hence $K$) as an adjustable parameter. For instance, partial ionisation from the 1s shell (in figure~\ref{fig:partial-K}) are better modelled with the formula eq.~\ref{eq:K-sc} from \citet[eq.(1)]{Guerra2012}. Furthermore, the $K$ parameter can be adjusted so as to match the BE$Q$ model to the experimental data in the peak region around 100 eV.
\end{enumerate} 

Finally, since there are no direct measurements of partial ionisation cross sections for electron impact ionisation, the further application of the BEB model for modelling partial ionisation processes will have to rely on three main sources of indirect comparisons:
\begin{enumerate}[1)]
	\item Total ionisation cross sections: further improvement will require including information about the differential dipole oscillator strengths integrated over band regions in the photoelectron spectrum for each transition considered, and therefrom using the BE$Q$ model instead of the BEB (as in \citet[§11.5.4]{Schmalzried2023}).
	\item Partial ionisation cross sections: use predictions published from sophisticated \textit{ab initio} models such as the B-spline R-matrix \citep{Zatsarinny2013} to assess the BE$Q$ partial cross sections.  
	\item Clastic ionisation cross sections: construct models or an average value of branching ratios \citep{Huber2019} for processes subsequent to direct ionisation (predissociation, autoionisation and radiative deexcitation) of molecular targets (production of ionic fragments).
\end{enumerate}

%
%
\subsection*{Acknowledgements}

The authors are grateful to the reviewers for their useful review and suggestions on how to improve the focus of the present article. We benefited from four different reviewers and we believe that their critique has led us to a very fruitful investigation. We acknowledge funding from the State Agency for Research of the Spanish MCIU through the ``Center of Excellence Severo Ochoa'' award for the Instituto de Astrofísica de Andalucía (SEV-2023-0709), and the Research Council of Norway under contracts 208028/F50, 216872/F50 and 223252/F50 (CoE). N. Lehtinen was supported by the Research Council of Norway under contract 335162/L40.

All the data gathered in this research are freely available through the \href{https://codeberg.org/aschmalz/elmolcs.git}{\texttt{elmolcs}} Python package hosted on Codeberg and now also on PyPI. The python file (\texttt{BEB.py}) used to produce all our graphs and our table is provided as a supplementary material.

\sisetup{round-mode=places,round-precision=2,round-pad=false}
\begin{longtable}{lS[table-column-width=1.5cm]S[table-column-width=1.5cm]cc}
	
		\caption{Input parameters to the BEB model in eq.~\ref{eq:beb}. Binding electron energies in orbitals identified as partial ionisation thresholds ($I$), kinetic scaling energies ($K$) and orbital occupation number or participation to the partial ionisation ($N$). Processes marked with a barred arrow ($\nrightarrow$) are optically spin-forbidden (require electron exchange). Dashed arrows ($\dashrightarrow$) symbolise that the final state may not be seen as a hole in a subshell but is rather a mixture. Arrows with question marks ($\stackrel{?}{\rightarrow}$) mean that the process has various subsequent branching ratios. Asterisks (*) mark the presence of predissociation. The last column gives the references for the processes identified and numbers taken. The nomenclature of molecular orbitals can be read in \citet[§4.3:p.253--259]{AMO2}} \label{tab:long} \\
		\endfirsthead
		\br
		Subshell or & {$I$} & {$K$} & $N$ & Reference \\ 
		Reaction & {(eV)} & {(eV)} & & \\
		\mr
		$(1s)^{-1} \stackrel{?}{\rightarrow} \text{C}^{++} + \mathrm{e}^-$ & 296.24 & 160 &  2 & \citep[table. IV:line 26]{Bruch1985} \\
		$(2s)^{-1} \rightarrow \text{C}^{+}\,^{2}\,{P}$ & 24.98 &  66.9 &  2/3 & \citep{NIST-ADS} \\
		$(2s)^{-1} \nrightarrow \text{C}^{+}\,^{2}\,{S}$ & 23.22 &  65.1 &  -- & \\
		$(2s)^{-1} \nrightarrow \text{C}^{+}\,^{2}\,{D}$ & 20.55 &  62.5 &  -- & \\
		$(2s)^{-1} \rightarrow \text{C}^{+}\,^{4}\,{P}$ & 16.59 &  58.5 &  4/3 & \\
		$(2p)^{-1} \rightarrow \text{C}^{+}\,^{2}\,{P}$ & 11.26 &  45.4 &  2 & \\
		\mr
		$(1s)^{-1} \stackrel{?}{\rightarrow} \text{N}^{++} + \mathrm{e}^-$ & 409.64 & 223 &  2 & \citep{SantAnna2011} \\
		$(2s)^{-1}\nrightarrow \text{N}^+\,^{1}\,{P}^o$ & 35.22 & 100.9 & -- & \citep{NIST-ADS} \\
		$(2s)^{-1}\rightarrow \text{N}^+\,^{3}\,{S}^o$ & 33.78 & 99.4 & $3/4$ & \\
		$(2s)^{-1}\nrightarrow \text{N}^+\,^{1}\,{D}^o$ & 31.42 & 97.1 & -- & \\
		$(2s)^{-1}\nrightarrow \text{N}^+\,^{3}\,{P}^o$ &  28.08 & 93.7 & -- &  \\
		$(2s)^{-1}\nrightarrow \text{N}^+\,^{3}\,{D}^o$ & 25.97 & 91.6 & -- & \\		
		$(2s)^{-1}\rightarrow \text{N}^+\,^{5}\,{S}^o$ &  20.34 & 86.0 & $5/4$ & \\
		$(2p)^{-1}\nrightarrow \text{N}^+\,^{1}\,{S}$ & 18.59 & 69.6 & --  & \\ 
		$(2p)^{-1}\nrightarrow \text{N}^+\,{}^{1}\,{D}$ & 16.43 & 67.5 & --  & \\ 
		$(2p)^{-1}\rightarrow \text{N}^+\,^{3}\,{P}$ & 14.545 & 65.6 & $3$ & \\
		
		\mr
		$(1s)^{-1} \stackrel{?}{\rightarrow} \text{O}^{++} + \mathrm{e}^-$  & 544.54 &  297 &  2 & \citep[§4.1]{Frati2020,Gorczyca2013}  \\
		$(2s)^{-1} \rightarrow \text{O}^{+}\, ^{2}P$  & 39.978 & 124.7 &  2/3 &  \citep{NIST-ADS} \\
		$(2s)^{-1} \nrightarrow \text{O}^{+}\, ^{2}S$  & 37.883 &  122.6 &  -- &  \\
		$(2s)^{-1} \nrightarrow \text{O}^{+}\, ^{2}D$  & 34.198 &  119.0 &  -- &  \\
		$(2s)^{-1} \rightarrow \text{O}^{+}\, ^{4}P$  & 28.496 &  113.3 &  4/3 &  \\
		$(2p)^{-1} \rightarrow \text{O}^{+}\, ^{2}P^o$  & 18.635 & 87.1  &  6/5 &  \\		
		$(2p)^{-1} \rightarrow \text{O}^{+}\, ^{2}D^o$  & 16.943 & 85.4 &  2 &  \\
		$(2p)^{-1} \rightarrow \text{O}^{+}\, ^{4}S^o$  & 13.618 & 82.8  &  4/5 &  \\		
		\mr
		$(1s)^{-1} \stackrel{?}{\rightarrow} \text{F}^{++} + \mathrm{e}^- $  & 696.7 &  381.3 &  2 & \citep[p.466]{Jolly1984}  \\
		$(2s)^{-1} \rightarrow \text{F}^{+}\, ^{1}P^o$  & 47.12  & 158.6  &  0.5 & \citep{NIST-ADS} \\
		$(2s)^{-1} \rightarrow \text{F}^{+}\, ^{3}P^o$  & 37.89 &  149.4 &  1.5 &  \\
		$(2p)^{-1} \rightarrow \text{F}^{+}\, ^{1}S$  & 22.99 & 113.9  &  1/3 &  \\		
		$(2p)^{-1} \rightarrow \text{F}^{+}\, ^{1}D$  & 20.01 & 110.9 &  5/3 &  \\
		$(2p)^{-1} \rightarrow \text{F}^{+}\, ^{3}P$  & 17.423 & 108.3  &  3 &  \\		
		\br
		$(1\sigma)^{-1} \stackrel{?}{\rightarrow} \text{C}^{++}, \text{O}^{++}, \text{C}^+ + \text{O}^+$ & 542.57 & 297 & 2 & \citep[p.170]{CDM} \\
		$(2\sigma)^{-1} \stackrel{?}{\rightarrow} \text{C}^{++}, \text{O}^{++}, \text{C}^+ + \text{O}^+$ & 296.24 & 160 & 2 & \\
		$(3\sigma)^{-1} \stackrel{?}{\rightarrow} \text{C}^+ + \text{O}$ & 38.9  & 120.4 & 2 & \\
		$(4\sigma)^{-1} \rightarrow \text{CO}^+\,\text{B}\,^{2}\Sigma^+$ & 19.7  & 92.9 & 2 & \\
		$(1\pi)^{-1} \rightarrow \text{CO}^+ \, \text{A}\,^{2}\Pi$ & 16.58 & 70.9 & 4 & \\
		$(5\sigma)^{-1} \rightarrow \text{CO}^+ \, \text{X}\,^{2}\Sigma^+$ & 14.01 & 56.3 & 2 & \\
		\mr
		$(1\sigma)^{-1} \stackrel{?}{\rightarrow} \text{N}_2^{++}, \text{N}^+ +\text{N}^+$ & 409.5 & 223 & 4 & \citep[p.426]{CDM} \\
		$(2\sigma_g)^{-1} \rightarrow \text{N}+\text{N}^+$ & 37.8  & 108.9 & 2 & \\
		$(2\sigma_u)^{-1} \dashrightarrow \text{N}_2^+ \, \text{C}\,^{2}\Sigma_u^+$ * & 23.6 & 86.8 & 1 &  \\
		$(2\sigma_u)^{-1} \dashrightarrow \text{N}_2^+ \, \text{B}\,^{2}\Sigma_u^+$ & 18.746 & 81.9 & 1 & \\
		$(1\pi_u)^{-1} \rightarrow \text{N}_2^+ \, \text{A}\,^{2}\Pi_u$ & 16.716 & 61.0 & 4 & \\
		$(3\sigma_g)^{-1}\rightarrow \text{N}_2^+ \, \text{X}\,^{2}\Sigma_g^+$ & 15.581 & 70.5 & 2 & \\
		\mr
		$(1\sigma)^{-1} \stackrel{?}{\rightarrow} \text{N}^+ + \text{O}^+,\, \text{N}^{++}, \, \text{O}^{++}$ & 543.2 & 297 & 2 & \citep[p.482]{CDM} \\
		$(2\sigma)^{-1} \stackrel{?}{\rightarrow} \text{N}^+ + \text{O}^+,\, \text{N}^{++}, \, \text{O}^{++}$ & 410.2 & 223 & 2 & \\
		$(3\sigma)^{-1} \stackrel{?}{\rightarrow} \text{N}^+ + \text{O} \, $  & 40.56  & 117.1 & 2 & \\		
		$(4\sigma)^{-1} \dashrightarrow \text{NO}^+ \, \text{c}\,^{3}\Pi,\,\text{B}\,^{1}\Pi$ * & 21.72  & 98.8 & 2 & \citep{Edqvist1971} \\		
		$(5\sigma)^{-1} \rightarrow \text{NO}^+\,\text{A}\,^{1}\Pi$ & 18.37  & 80.6 & 0.5 &  \\
		$(5\sigma)^{-1} \rightarrow \text{NO}^+ \, \text{b}\,^{3}\Pi$ & 16.6 & 78.9 & 1.5 & \\
		$(1\pi)^{-1} \stackrel{?}{\rightarrow} \text{NO}^+\, \text{B}'\,^{1}\Sigma^+$ & 22.727  & 78.1 & 1/3 & \\
		$(1\pi)^{-1} \rightarrow \text{NO}^+ \,\text{W}\,^{1}\Delta$ & 18.12  & 73.5 & 1/3 & \\
		$(1\pi)^{-1} \rightarrow \text{NO}^+ \, \text{A}'\,^{1}\Sigma^+$ & 17.88  & 73.3 & 1/3 & \\
		$(1\pi)^{-1} \rightarrow \text{NO}^+ \, \text{b}'\,^{3}\Sigma^-$ & 17.65  & 73.0 & 1 & \\		
		$(1\pi)^{-1} \rightarrow \text{NO}^+\,\text{w}\,^{3}\Delta$ & 16.93  & 72.3 & 1 & \\
		$(1\pi)^{-1} \rightarrow \text{NO}^+ \, \text{a}\,^{3}\Sigma^+$& 15.73 & 71.1 & 1 & \\
		$(2\pi)^{-1} \rightarrow \text{NO}^+ \, \text{X}\,^{1}\Sigma^+$ & 9.26 & 74.5 & 1 & \\		
		\mr
		$(1\sigma)^{-1} \stackrel{?}{\rightarrow} \text{O}_2^+, \text{O}_2^{++}, \text{O}+\text{O}^+$ & 543.1 & 297 & 4 & \citep[p.504]{CDM} \\
		$(2\sigma_g)^{-1} \stackrel{?}{\rightarrow} \text{O}_2^+, \text{O}+\text{O}^+$ & 39.57  & 79.73 & 2 &  \\
		$(2\sigma_u)^{-1} \rightarrow \text{O}_2^+\,\text{c}\,^{4}\Sigma_u^-$ * & 24.58  & 90.92 & 2 & \citep[p.222--5]{Berkowitz1979} \\
		$(3\sigma_g)^{-1} \rightarrow \text{O}_2^+ \, \text{B}\,^{2}\Sigma_g^-$ * & 20.3 & 71.84 & 2/3 & \citep{Edqvist1970} \\
		$(3\sigma_g)^{-1} \rightarrow \text{O}_2^+ \, \text{b}\,^{4}\Sigma_g^-$ * & 18.17 & 71.84 & 4/3 & \\
		$(1\pi_u)^{-1}\rightarrow \text{O}_2^+ \, \text{A}\,^{2}\Pi_u$ & 17.15 & 59.89 & 4/3 & \\		
		$(1\pi_u)^{-1} \rightarrow \text{O}_2^+ \, \text{a}\,^{4}\Pi_u$ & 16.157 & 59.89 & 8/3 & \\
		$(1\pi_g)^{-1}\rightarrow \text{O}_2^+ \, \text{X}\,^{2}\Pi_g$ & 12.07 & 84.88 & 2 & \\	
		\br
		$(1\sigma)^{-1} \stackrel{?}{\rightarrow} \text{C}^+ 2 + \text{O}^+$ & 541.1 & 297 & 4 &  \citep[tab.6:p.147]{Allan1972} \\ 
		$(2\sigma_g)^{-1} \stackrel{?}{\rightarrow} \text{C}^+ 2 + \text{O}^+$ & 297.5 & 160 & 2 &  \\
		$(2\sigma_u)^{-1} \stackrel{?}{\rightarrow} \text{C}^+ + 2\text{O}$ & 39.7 & 115.4 & 2 & \citep{Aresta2016} \\
		$(3\sigma_g)^{-1} \stackrel{?}{\rightarrow} \text{C}^+ + 2\text{O}$ & 37.5 & 115.9 & 2 &  \\
		$(4\sigma_g)^{-1} \rightarrow \text{CO}_2^+ \, \tilde{\text{C}}\,{}^{2}\Sigma_g^+ *$ & 19.4 & 94.7 & 2 &  \\
		$(3\sigma_u)^{-1} \rightarrow \text{CO}_2^+ \, \tilde{\text{B}}\,^{2}\Sigma_u^+$ & 18.1 & 89.7 & 2 &  \\
		$(1\pi_u)^{-1} \rightarrow \text{CO}_2^+ \, \tilde{\text{A}}\,^{2}\Pi_u$ & 17.6 & 67.6 & 4 &  \\
		$(1\pi_g)^{-1} \rightarrow \text{CO}_2^+ \, \text{X}\,^{2}\Pi_g$ & 13.8 & 78.2 & 4 &  \\
		\mr
		$(1\text{a}_1)^{-1} \stackrel{?}{\rightarrow} 2\text{H}^+ + \text{O}^+$ & 539.70 & 297 & 2 & \citep[p.84:table~5.3.1]{ESCA2} \\
		$(2\text{a}_1)^{-1} \stackrel{?}{\rightarrow} \text{H}^+ + \text{H} + \text{O}$ & 32.20 & 102.9 & 2 &  \\
		$(1\text{b}_2)^{-1} \rightarrow \text{H}_2\text{O}^{+}\,{}^2\text{B}_2$ & 18.51 & 66.87 & 2 & \citep[p.241:table~5]{Berkowitz1979} \\
		$(3\text{a}_1)^{-1} \rightarrow \text{H}_2\text{O}^{+}\,{}^2\text{A}_1$ & 14.74 & 74.27 & 2 &  \\
		$(1\text{b}_1)^{-1} \rightarrow \text{H}_2\text{O}^{+}\,{}^2\text{B}_1$ & 12.62 & 74.52 & 2 &  \\
		\mr
		$(1\text{a}_1)^{-1} \stackrel{?}{\rightarrow} \text{N} + 2\text{O}^+$ & 541.3 & 297 & 4 & \citep[p.126:table 5.4.4]{ESCA2} \\
		$(2\text{a}_1)^{-1} \stackrel{?}{\rightarrow} \text{N}^+ + \text{O}^+ + \text{O}$ & 412.4 & 223 & 2 &  \citep[p.123:table 5.4.3]{ESCA2} \\
		$(3\text{a}_1)^{-1} \stackrel{?}{\rightarrow} \text{NO}^+ + \text{O}$ & 38.9 & 112 & 2 & \citep[table~2]{Baltzer1998} \\
		$(2\text{b}_2)^{-1} \stackrel{?}{\rightarrow} \text{NO}^+ + \text{O}$ & 32.0 & 110 & 2 &  \\
		$(4\text{a}_1)^{-1} \stackrel{?}{\rightarrow} \text{NO}^+ + \text{O}$ & 21.3 & 100 & 2 &  \\
		$(3\text{b}_2)^{-1} \stackrel{?}{\rightarrow} \text{NO} + \text{O}^+$ & 19.06 & 75.8 & 2 &  \\
		$(1\text{b}_1)^{-1} \dashrightarrow \text{NO}_2^{+}\,\text{d}\,{}^3\text{B}_1$* & 17.63 & 90.9 & 2 &  \\
		$(5\text{a}_1)^{-1} \rightarrow \text{NO}_2^{+}\,\text{c}\,{}^3\text{A}_1$* & 17.45 & 69 & 2 &  \\
		$(4\text{b}_2)^{-1} \rightarrow \text{NO}_2^{+}\,\text{B}\,{}^1\text{B}_2$* & 14.60 & 83.7 & 0.5 &  \\
		$(1\text{a}_2)^{-1} \rightarrow \text{NO}_2^{+}\,\text{A}\,{}^1\text{A}_2$* & 14.13 & 83.2 & 0.5 &  \\
		$(1\text{a}_2)^{-1} \rightarrow \text{NO}_2^{+}\,\text{b}\,{}^3\text{A}_2$* & 13.69 & 78.9 & 1.5 &  \\
		$(4\text{b}_2)^{-1} \rightarrow \text{NO}_2^{+}\,\text{a}\,{}^3\text{B}_2$* & 13.06 & 78.3 & 1.5 &  \\
		$(6\text{a}_1)^{-1} \rightarrow \text{NO}_2^{+}\,\text{X}\,{}^1\Sigma_g^+$ & 11.2 & 87.4 & 1 &  \\
		\mr
		$(1\text{a}_1)^{-1} \stackrel{?}{\rightarrow} 2\text{O}^+ + \text{O}$ & 546.20 & 297 & 2 & \citep{Banna1977} \\
		$(2\text{a}_1)^{-1},\,(1\text{b}_2)^{-1} \stackrel{?}{\rightarrow} 2\text{O}^+ + \text{O}$ & 541.50 & 297 & 4 &  \\
		$[(3\text{a}_1)^{-1}] \stackrel{?}{\dashrightarrow} \text{O}_2^+ + \text{O}$ & 50.16 & 128.37 & 2 & \citep[table~II]{Kim1997} \\
		$[(2\text{b}_2)^{-1}] \stackrel{?}{\dashrightarrow} \text{O}_2^+ + \text{O}$ & 40.63 & 121.25 & 2 &  \\
		$[(4\text{a}_1)^{-1}] \stackrel{?}{\dashrightarrow} \text{O}_2^+ + \text{O}$ & 29.11 & 118.45 & 2 &  \\
		$ [\text{O}_3^{+}\,6\,{}^2\text{B}_2, \;5\,{}^2\text{B}_1,\; \ldots $]* & 20.00 & 89.13 & [2] & \citep[table~I]{Couto2006} \\
		$ [\text{O}_3^{+}\,3\,{}^2\text{A}_1, \;4\,{}^2\text{B}_2 $]* & 17.30 & 75.85 & [2] &  \\
		$ [\text{O}_3^{+}\,2\,{}^2\text{B}_1$]* & 16.57 & 93.66 & [2] &  \\
		$(1\text{a}_2)^{-1} \dashrightarrow \text{O}_3^{+}\,\tilde{\text{B}}{}^2\text{A}_2$* & 13.54 & 87.36 & 2 &  \\
		$(4\text{b}_2)^{-1} \dashrightarrow \text{O}_3^{+}\,\tilde{\text{A}}{}^2\text{B}_2$ & 13.00 & 89.56 & 2 &  \\
		$(6\text{a}_1)^{-1} \dashrightarrow \text{O}_3^{+}\,\tilde{\text{X}}{}^2\text{A}_1$ & 12.73 & 77.85 & 2 &  \\
		\br
\end{longtable}

\bibliographystyle{plainnat}
\bibliography{references}

\begin{thebibliography}{94}
\providecommand{\natexlab}[1]{#1}
\providecommand{\url}[1]{\texttt{#1}}
\expandafter\ifx\csname urlstyle\endcsname\relax
  \providecommand{\doi}[1]{doi: #1}\else
  \providecommand{\doi}{doi: \begingroup \urlstyle{rm}\Url}\fi

\bibitem[Allan et~al.(1972)Allan, Gelius, Allison, Johansson, Siegbahn, and
  Siegbahn]{Allan1972}
C.J. Allan, U.~Gelius, D.A. Allison, G.~Johansson, H.~Siegbahn, and
  K.~Siegbahn.
\newblock {ESCA studies of CO2, CS2 and COS}.
\newblock \emph{J. Electron Spectrosc. Relat. Phenom.}, 1\penalty0
  (2):\penalty0 131--151, 1972.
\newblock ISSN 0368-2048.
\newblock \doi{https://doi.org/10.1016/0368-2048(72)80027-6}.
\newblock URL
  \url{https://www.sciencedirect.com/science/article/pii/0368204872800276}.

\bibitem[Almeida et~al.(1995)Almeida, Fontes, and Godinho]{Almeida1995}
D~P Almeida, A~C Fontes, and C~F~L Godinho.
\newblock Electron-impact ionization cross section of neon ( sigma n+, n=1-5).
\newblock \emph{J. Phys. B: At. Mol. Opt. Phys.}, 28\penalty0 (15):\penalty0
  3335, aug 1995.
\newblock \doi{10.1088/0953-4075/28/15/022}.
\newblock URL \url{https://dx.doi.org/10.1088/0953-4075/28/15/022}.

\bibitem[Aresta et~al.(2016)Aresta, Dibenedetto, and Quaranta]{Aresta2016}
Michele Aresta, Angela Dibenedetto, and Eugenio Quaranta.
\newblock \emph{The Carbon Dioxide Molecule}, pages 1--34.
\newblock Springer Berlin Heidelberg, Berlin, Heidelberg, 2016.
\newblock ISBN 978-3-662-46831-9.
\newblock \doi{10.1007/978-3-662-46831-9_1}.
\newblock URL \url{https://doi.org/10.1007/978-3-662-46831-9_1}.

\bibitem[Au and Brion(1997)]{Au1997}
Jennifer~W. Au and C.~E. Brion.
\newblock Absolute oscillator strenghts for the valence-shell photoabsorption
  {(2–200 eV)} and the molecular and dissociative photoionization {(11–80
  eV)} of nitrogen dioxide.
\newblock \emph{Chem. Phys.}, 218\penalty0 (1):\penalty0 109--126, 1997.
\newblock ISSN 0301-0104.
\newblock \doi{https://doi.org/10.1016/S0301-0104(97)00065-7}.
\newblock URL
  \url{https://www.sciencedirect.com/science/article/pii/S0301010497000657}.

\bibitem[Baltzer et~al.(1998)Baltzer, Karlsson, Wannberg, Holland, MacDonald,
  Hayes, and Eland]{Baltzer1998}
P.~Baltzer, L.~Karlsson, B.~Wannberg, D.M.P. Holland, M.A. MacDonald, M.A.
  Hayes, and J.H.D. Eland.
\newblock An experimental study of the valence shell photoelectron spectrum of
  the {NO2} molecule.
\newblock \emph{Chem. Phys.}, 237\penalty0 (3):\penalty0 451--470, 1998.
\newblock ISSN 0301-0104.
\newblock \doi{https://doi.org/10.1016/S0301-0104(98)00240-7}.
\newblock URL
  \url{https://www.sciencedirect.com/science/article/pii/S0301010498002407}.

\bibitem[Banna et~al.(1977)Banna, Frost, McDowell, Noodleman, and
  Wallbank]{Banna1977}
M.Salim Banna, David~C. Frost, Charles~A. McDowell, Louis Noodleman, and Barry
  Wallbank.
\newblock A study of the core electron binding energies of ozone by x-ray
  photoelectron spectroscopy and the {X$\alpha$} scattered wave method.
\newblock \emph{Chem. Phys. Lett.}, 49\penalty0 (2):\penalty0 213--217, 1977.
\newblock ISSN 0009-2614.
\newblock \doi{https://doi.org/10.1016/0009-2614(77)80572-1}.
\newblock URL
  \url{https://www.sciencedirect.com/science/article/pii/0009261477805721}.

\bibitem[Berkowitz(1979)]{Berkowitz1979}
Joseph Berkowitz.
\newblock Chapter {VI} - partial cross sections.
\newblock In Joseph Berkowitz, editor, \emph{Photoabsorption, Photoionization,
  and Photoelectron Spectroscopy}, pages 155--357. Academic Press, 1979.
\newblock ISBN 978-0-12-091650-4.
\newblock \doi{https://doi.org/10.1016/B978-0-12-091650-4.50012-8}.
\newblock URL
  \url{https://www.sciencedirect.com/science/article/pii/B9780120916504500128}.

\bibitem[Brook et~al.(1978)Brook, Harrison, and Smith]{Brook1978}
E~Brook, M~F~A Harrison, and A~C~H Smith.
\newblock Measurements of the electron impact ionisation cross sections of {He,
  C, O and N} atoms.
\newblock \emph{J. Phys. B: At. Mol. Phys.}, 11\penalty0 (17):\penalty0
  3115--3132, September 1978.
\newblock \doi{10.1088/0022-3700/11/17/021}.
\newblock URL \url{https://doi.org/10.1088/0022-3700/11/17/021}.

\bibitem[Bruch et~al.(1985)Bruch, Chung, Luken, and Culberson]{Bruch1985}
Reinhard Bruch, Kwong~T. Chung, William~L. Luken, and John~C. Culberson.
\newblock Recalibration of the {KLL Auger} spectrum of carbon.
\newblock \emph{Phys. Rev. A}, 31:\penalty0 310--315, Jan 1985.
\newblock \doi{10.1103/PhysRevA.31.310}.
\newblock URL \url{https://link.aps.org/doi/10.1103/PhysRevA.31.310}.

\bibitem[Cade et~al.(1966)Cade, Sales, and Wahl]{Cade1966}
Paul~E. Cade, K.~D. Sales, and Arnold~C. Wahl.
\newblock Electronic structure of diatomic molecules. {III. A. Hartree—Fock}
  wavefunctions and energy quantities for {N$_2$(X${}^1\Sigma_g^+$) and
  N$_2^+$(X${}^2\Sigma_g^+$, A${}^2\Pi_u$, B${}^2\Sigma_u^+$)} molecular ions.
\newblock \emph{J. Chem. Phys.}, 44\penalty0 (5):\penalty0 1973--2003, 03 1966.
\newblock ISSN 0021-9606.
\newblock \doi{10.1063/1.1726972}.
\newblock URL \url{https://doi.org/10.1063/1.1726972}.

\bibitem[Chang and Huang(2009)]{Chang2009}
Hai-Bo Chang and Ming-Bao Huang.
\newblock A theoretical study on the electronic states and {O}-loss
  photodissociation of the {NO2+} ion.
\newblock \emph{ChemPhysChem}, 10\penalty0 (3):\penalty0 582--589, 2009.
\newblock \doi{https://doi.org/10.1002/cphc.200800626}.
\newblock URL
  \url{https://chemistry-europe.onlinelibrary.wiley.com/doi/abs/10.1002/cphc.200800626}.

\bibitem[Chanrion and Neubert(2010)]{Chanrion2010}
O.~Chanrion and T.~Neubert.
\newblock Production of runaway electrons by negative streamer discharges.
\newblock \emph{J. Geophys. Res. : Space Phys.}, 115\penalty0 (A00E32), 2010.
\newblock \doi{10.1029/2009JA014774}.
\newblock URL
  \url{https://agupubs.onlinelibrary.wiley.com/doi/abs/10.1029/2009JA014774}.

\bibitem[Clementi and Raimondi(1963)]{Clementi1963}
E.~Clementi and D.~L. Raimondi.
\newblock {Atomic Screening Constants from SCF Functions}.
\newblock \emph{J. Chem. Phys.}, 38\penalty0 (11):\penalty0 2686--2689, 06
  1963.
\newblock ISSN 0021-9606.
\newblock \doi{10.1063/1.1733573}.
\newblock URL \url{https://doi.org/10.1063/1.1733573}.

\bibitem[Clemmer and Armentrout(1992)]{Clemmer1992}
D.~E. Clemmer and P.~B. Armentrout.
\newblock {Direct determination of the adiabatic ionization energy of NO2 as
  measured by guided ion‐beam mass spectrometry}.
\newblock \emph{J. Chem. Phys.}, 97\penalty0 (4):\penalty0 2451--2458, 08 1992.
\newblock ISSN 0021-9606.
\newblock \doi{10.1063/1.463083}.
\newblock URL \url{https://doi.org/10.1063/1.463083}.

\bibitem[Couto et~al.(2006)Couto, Mocellin, Moreira, Gomes, Naves~de Brito, and
  Lopes]{Couto2006}
H.~Couto, A.~Mocellin, C.~D. Moreira, M.~P. Gomes, A.~Naves~de Brito, and
  M.~C.~A. Lopes.
\newblock {Threshold photoelectron spectroscopy of ozone}.
\newblock \emph{J. Chem. Phys.}, 124\penalty0 (20):\penalty0 204311, 05 2006.
\newblock ISSN 0021-9606.
\newblock \doi{10.1063/1.2200702}.
\newblock URL \url{https://doi.org/10.1063/1.2200702}.

\bibitem[Célestin and Pasko(2011)]{Celestin2011}
Sébastien Célestin and Victor~P. Pasko.
\newblock Energy and fluxes of thermal runaway electrons produced by
  exponential growth of streamers during the stepping of lightning leaders and
  in transient luminous events.
\newblock \emph{J. Geophys. Res. : Space Phys.}, 116\penalty0 (A3), 2011.
\newblock \doi{10.1029/2010JA016260}.
\newblock URL
  \url{https://agupubs.onlinelibrary.wiley.com/doi/abs/10.1029/2010JA016260}.

\bibitem[Deutsch et~al.(2004)Deutsch, Scheier, Becker, and Märk]{Deutsch2004}
H.~Deutsch, P.~Scheier, K.~Becker, and T.~D. Märk.
\newblock Revised high energy behavior of the {Deutsch-Märk (DM)} formula for
  the calculation of electron impact ionization cross sections of atoms.
\newblock \emph{Int J Mass Spectrom}, 233\penalty0 (1):\penalty0 13--17, 2004.
\newblock ISSN 1387-3806.
\newblock \doi{https://doi.org/10.1016/j.ijms.2003.06.002}.
\newblock URL
  \url{https://www.sciencedirect.com/science/article/pii/S1387380603004962}.
\newblock Special Issue: In honour of Tilmann Mark.

\bibitem[Djurić et~al.(1988)Djurić, Čadež, and Kurepa]{Djuric1988}
N.Lj. Djurić, I.M. Čadež, and M.V. Kurepa.
\newblock {H2O} and {D2O} total ionization cross-sections by electron impact.
\newblock \emph{Int. J. Mass Spectrom.}, 83\penalty0 (3):\penalty0 R7--R10,
  1988.
\newblock ISSN 0168-1176.
\newblock \doi{https://doi.org/10.1016/0168-1176(88)80038-7}.
\newblock URL
  \url{https://www.sciencedirect.com/science/article/pii/0168117688800387}.

\bibitem[Duke and O'Leary(1995)]{Duke1995}
Brian~J. Duke and Brian O'Leary.
\newblock {Non-Koopmans' Molecules}.
\newblock \emph{J. Chem. Educ.}, 72\penalty0 (6):\penalty0 501, June 1995.
\newblock ISSN 0021-9584.
\newblock \doi{10.1021/ed072p501}.
\newblock URL \url{https://doi.org/10.1021/ed072p501}.

\bibitem[Edqvist et~al.(1970)Edqvist, Lindholm, Selin, and
  {\AA}sbrink]{Edqvist1970}
O.~Edqvist, E.~Lindholm, L.~E. Selin, and L.~{\AA}sbrink.
\newblock On the photoelectron spectrum of {O$_2$}.
\newblock \emph{Phys Scripta}, 1\penalty0 (1):\penalty0 25--30, January 1970.
\newblock \doi{10.1088/0031-8949/1/1/004}.
\newblock URL \url{https://doi.org/10.1088/0031-8949/1/1/004}.

\bibitem[Edqvist et~al.(1971)Edqvist, åsbrink, and Lindholm]{Edqvist1971}
O.~Edqvist, L.~åsbrink, and E.~Lindholm.
\newblock On the photoelectron spectrum of {NO}.
\newblock \emph{Zeitschrift für Naturforschung A}, 26\penalty0 (9):\penalty0
  1407--1410, 1971.
\newblock \doi{doi:10.1515/zna-1971-0906}.
\newblock URL \url{https://doi.org/10.1515/zna-1971-0906}.

\bibitem[Eland and Karlsson(1998)]{Eland1998}
J.H.D. Eland and L.~Karlsson.
\newblock Dissociative photoionisation of {NO2} up to {26 eV}.
\newblock \emph{Chem. Phys.}, 237\penalty0 (1):\penalty0 139--148, 1998.
\newblock ISSN 0301-0104.
\newblock \doi{https://doi.org/10.1016/S0301-0104(98)00237-7}.
\newblock URL
  \url{https://www.sciencedirect.com/science/article/pii/S0301010498002377}.

\bibitem[Erman(1976)]{Erman1976}
Peter Erman.
\newblock Direct measurement of the {N2+ C} state predissociation probability.
\newblock \emph{Phys Scr}, 14\penalty0 (1-2):\penalty0 51, July 1976.
\newblock \doi{10.1088/0031-8949/14/1-2/010}.
\newblock URL \url{https://dx.doi.org/10.1088/0031-8949/14/1-2/010}.

\bibitem[Frati et~al.(2020)Frati, Hunault, and de~Groot]{Frati2020}
Federica Frati, Myrtille O. J.~Y. Hunault, and Frank M.~F. de~Groot.
\newblock Oxygen {K}-edge {X}-ray absorption spectra.
\newblock \emph{Chem. Rev.}, 120\penalty0 (9):\penalty0 4056--4110, 2020.
\newblock \doi{10.1021/acs.chemrev.9b00439}.
\newblock URL \url{https://doi.org/10.1021/acs.chemrev.9b00439}.
\newblock PMID: 32275144.

\bibitem[{Gaudin, Albert} and {Hagemann, Robert}(1967)]{Gaudin1967}
{Gaudin, Albert} and {Hagemann, Robert}.
\newblock D\'eterminations absolues des sections efficaces totales et
  partielles d\'{}ionisation de l\'{}h\'elium, du n\'eon, de l\'{}argon et de
  l\'{}ac\'etyl\`ene, pour des \'electrons de 100 \`a 2000 {eV}.
\newblock \emph{J. Chim. Phys.}, 64:\penalty0 1209--1221, 1967.
\newblock \doi{10.1051/jcp/1967641209}.
\newblock URL \url{https://doi.org/10.1051/jcp/1967641209}.

\bibitem[Gedeon et~al.(2014)Gedeon, Gedeon, Lazur, Nagy, Zatsarinny, and
  Bartschat]{Gedeon2014F}
Viktor Gedeon, Sergej Gedeon, Vladimir Lazur, Elizabeth Nagy, Oleg Zatsarinny,
  and Klaus Bartschat.
\newblock $b$-spline $r$-matrix-with-pseudostates calculations for
  electron-impact excitation and ionization of fluorine.
\newblock \emph{Phys. Rev. A}, 89:\penalty0 052713, May 2014.
\newblock \doi{10.1103/PhysRevA.89.052713}.
\newblock URL \url{https://link.aps.org/doi/10.1103/PhysRevA.89.052713}.

\bibitem[Gorczyca et~al.(2013)Gorczyca, Bautista, Hasoglu, García, Gatuzz,
  Kaastra, Kallman, Manson, Mendoza, Raassen, de~Vries, and
  Zatsarinny]{Gorczyca2013}
T.~W. Gorczyca, M.~A. Bautista, M.~F. Hasoglu, J.~García, E.~Gatuzz, J.~S.
  Kaastra, T.~R. Kallman, S.~T. Manson, C.~Mendoza, A.~J.~J. Raassen, C.~P.
  de~Vries, and O.~Zatsarinny.
\newblock A comprehensive {X}-ray absorption model for atomic oxygen.
\newblock \emph{ApJ}, 779\penalty0 (1):\penalty0 78, nov 2013.
\newblock \doi{10.1088/0004-637X/779/1/78}.
\newblock URL \url{https://dx.doi.org/10.1088/0004-637X/779/1/78}.

\bibitem[Guerra et~al.(2012)Guerra, Parente, Indelicato, and
  Santos]{Guerra2012}
M.~Guerra, F.~Parente, P.~Indelicato, and J.~P. Santos.
\newblock Modified binary encounter {B}ethe model for electron-impact
  ionization.
\newblock \emph{Int J Mass Spectrom}, 313:\penalty0 1--7, 2012.
\newblock ISSN 1387-3806.
\newblock \doi{https://doi.org/10.1016/j.ijms.2011.12.003}.
\newblock URL
  \url{https://www.sciencedirect.com/science/article/pii/S138738061100474X}.

\bibitem[Hagelaar and Pitchford(2005)]{Hagelaar2005}
G.~J.~M. Hagelaar and L.~C. Pitchford.
\newblock Solving the boltzmann equation to obtain electron transport
  coefficients and rate coefficients for fluid models.
\newblock \emph{Plasma Sources Sci. Technol.}, 14\penalty0 (4):\penalty0 722,
  October 2005.
\newblock \doi{10.1088/0963-0252/14/4/011}.
\newblock URL \url{https://dx.doi.org/10.1088/0963-0252/14/4/011}.

\bibitem[Hayes et~al.(1987)Hayes, Wetzel, and Freund]{Hayes1987}
Todd~R. Hayes, Robert~C. Wetzel, and Robert~S. Freund.
\newblock Absolute electron-impact-ionization cross-section measurements of the
  halogen atoms.
\newblock \emph{Phys. Rev. A}, 35:\penalty0 578--584, Jan 1987.
\newblock \doi{10.1103/PhysRevA.35.578}.
\newblock URL \url{https://link.aps.org/doi/10.1103/PhysRevA.35.578}.

\bibitem[Hertel and Schulz(2015)]{AMO2}
Ingolf~V. Hertel and Claus-Peter Schulz.
\newblock \emph{{Atoms, Molecules and Optical Physics 2: Molecules and Photons
  - Spectroscopy and Collisions}}.
\newblock Springer Berlin Heidelberg, Berlin, Heidelberg, 2015.
\newblock ISBN 978-3-642-54313-5.
\newblock \doi{10.1007/978-3-642-54313-5}.
\newblock URL \url{https://doi.org/10.1007/978-3-642-54313-5}.

\bibitem[Huber(1968)]{Huber1968}
K.~P. Huber.
\newblock The excited electronic states of the {NO+} ion.
\newblock \emph{Can J Phys}, 46\penalty0 (15):\penalty0 1691--1695, 1968.
\newblock \doi{10.1139/p68-501}.
\newblock URL \url{https://doi.org/10.1139/p68-501}.

\bibitem[Huber and Herzberg(1979)]{CDM}
K.~P. Huber and G.~Herzberg.
\newblock \emph{{IV. Constants of Diatomic Molecules}}.
\newblock Molecular Spectra and Molecular Structure. Springer US, Boston, MA,
  1979.
\newblock ISBN 978-1-4757-0961-2.
\newblock \doi{10.1007/978-1-4757-0961-2_2}.
\newblock URL \url{https://doi.org/10.1007/978-1-4757-0961-2_2}.

\bibitem[Huber et~al.(2019)Huber, Mauracher, Süß, Sukuba, Urban, Borodin, and
  Probst]{Huber2019}
Stefan~E. Huber, Andreas Mauracher, Daniel Süß, Ivan Sukuba, Jan Urban,
  Dmitry Borodin, and Michael Probst.
\newblock Total and partial electron impact ionization cross sections of
  fusion-relevant diatomic molecules.
\newblock \emph{J. Chem. Phys.}, 150\penalty0 (2):\penalty0 024306, 2019.
\newblock \doi{10.1063/1.5063767}.
\newblock URL \url{https://doi.org/10.1063/1.5063767}.

\bibitem[Huo and Kim(1999)]{Huo1999}
W.~M. Huo and Y.-K. Kim.
\newblock Electron collision cross-section data for plasma modeling.
\newblock \emph{IEEE Trans. Plasma Sci.}, 27\penalty0 (5):\penalty0 1225--1240,
  1999.
\newblock \doi{10.1109/27.799798}.

\bibitem[Hwang et~al.(1996)Hwang, Kim, and Rudd]{Hwang1996}
W.~Hwang, Y.‐K. Kim, and M.~E. Rudd.
\newblock New model for electron‐impact ionization cross sections of
  molecules.
\newblock \emph{J. Chem. Phys.}, 104\penalty0 (8):\penalty0 2956--2966, 1996.
\newblock \doi{10.1063/1.471116}.
\newblock URL \url{https://doi.org/10.1063/1.471116}.

\bibitem[Hönicke et~al.(2023)Hönicke, Unterumsberger, Wauschkuhn, Krämer,
  Beckhoff, Indelicato, Sampaio, Marques, Guerra, Parente, and
  Santos]{Honicke2023}
Philipp Hönicke, Rainer Unterumsberger, Nils Wauschkuhn, Markus Krämer,
  Burkhard Beckhoff, Paul Indelicato, Jorge Sampaio, José~Pires Marques, Mauro
  Guerra, Fernando Parente, and José~Paulo Santos.
\newblock Experimental and theoretical approaches for determining the {K}-shell
  fluorescence yield of carbon.
\newblock \emph{Radiat. Phys. Chem.}, 202:\penalty0 110501, 2023.
\newblock ISSN 0969-806X.
\newblock \doi{https://doi.org/10.1016/j.radphyschem.2022.110501}.
\newblock URL
  \url{https://www.sciencedirect.com/science/article/pii/S0969806X22005370}.

\bibitem[Inokuti et~al.(2000)Inokuti, Buckman, Elford, and Tawara]{phe-Atom-2}
M.~Inokuti, S.~J. Buckman, M.~T. Elford, and H.~Tawara.
\newblock 2. {E}lectron collisions with atoms.
\newblock In Y.~Itikawa, editor, \emph{Interactions of Photons and Electrons
  with Atoms}, volume 17 A of \emph{Landolt-B{\"o}rnstein - Group I Elementary
  Particles, Nuclei and Atoms}, chapter~2. Springer-Verlag Berlin Heidelberg,
  2000.
\newblock \doi{10.1007/10547143_2}.
\newblock URL
  \url{https://materials.springer.com/lb/docs/sm_lbs_978-3-540-69716-9_2}.

\bibitem[Inokuti(1971)]{Inokuti1971}
Mitio Inokuti.
\newblock Inelastic collisions of fast charged particles with atoms and
  molecules---{The Bethe} theory revisited.
\newblock \emph{Rev. Mod. Phys.}, 43:\penalty0 297--347, July 1971.
\newblock \doi{10.1103/RevModPhys.43.297}.
\newblock URL \url{https://link.aps.org/doi/10.1103/RevModPhys.43.297}.
\newblock In this work, $M$ represents the reduced mass of the system $m_1
  m_2/(m_1+m_2)$, which is noted $\mu$ in our work.

\bibitem[Itikawa(2006)]{Itikawa2006}
Yukikazu Itikawa.
\newblock Cross sections for electron collisions with nitrogen molecules.
\newblock \emph{J Phys Chem Ref Data}, 35\penalty0 (1):\penalty0 31--53, 2006.
\newblock \doi{10.1063/1.1937426}.
\newblock URL \url{https://doi.org/10.1063/1.1937426}.

\bibitem[Itikawa(2009)]{Itikawa2009}
Yukikazu Itikawa.
\newblock Cross sections for electron collisions with oxygen molecules.
\newblock \emph{J Phys Chem Ref Data}, 38\penalty0 (1):\penalty0 1--20, 2009.
\newblock \doi{10.1063/1.3025886}.
\newblock URL \url{https://doi.org/10.1063/1.3025886}.

\bibitem[Itikawa(2016)]{Itikawa2016}
Yukikazu Itikawa.
\newblock Cross sections for electron collisions with nitric oxide.
\newblock \emph{J Phys Chem Ref Data}, 45\penalty0 (3):\penalty0 033106, 2016.
\newblock \doi{10.1063/1.4961372}.
\newblock URL \url{https://doi.org/10.1063/1.4961372}.

\bibitem[Jolly et~al.(1984)Jolly, Bomben, and Eyermann]{Jolly1984}
W.~L. Jolly, K.~D. Bomben, and C.~J. Eyermann.
\newblock Core-electron binding energies for gaseous atoms and molecules.
\newblock \emph{Atom Data Nucl Data}, 31\penalty0 (3):\penalty0 433--493, 1984.
\newblock ISSN 0092-640X.
\newblock \doi{https://doi.org/10.1016/0092-640X(84)90011-1}.
\newblock URL
  \url{http://www.sciencedirect.com/science/article/pii/0092640X84900111}.

\bibitem[Karlsson et~al.(1975)Karlsson, Mattsson, Jadrny, Albridge, Pinchas,
  Bergmark, and Siegbahn]{Karlsson1975}
L.~Karlsson, L.~Mattsson, R.~Jadrny, R.~G. Albridge, S.~Pinchas, T.~Bergmark,
  and K.~Siegbahn.
\newblock {Isotopic and vibronic coupling effects in the valence electron
  spectra of H2~16O, H2~18O, and D2~16O}.
\newblock \emph{J. Chem. Phys.}, 62\penalty0 (12):\penalty0 4745--4752, 06
  1975.
\newblock ISSN 0021-9606.
\newblock \doi{10.1063/1.430423}.
\newblock URL \url{https://doi.org/10.1063/1.430423}.

\bibitem[Kim et~al.(1981)Kim, Stephan, Märk, and Märk]{Kim1981}
Y.~B. Kim, K.~Stephan, E.~Märk, and T.~D. Märk.
\newblock {Single and double ionization of nitric oxide by electron impact from
  threshold up to 180 eV}.
\newblock \emph{J. Chem. Phys.}, 74\penalty0 (12):\penalty0 6771--6776, 06
  1981.
\newblock ISSN 0021-9606.
\newblock \doi{10.1063/1.441082}.
\newblock URL \url{https://doi.org/10.1063/1.441082}.

\bibitem[Kim et~al.(1997)Kim, Hwang, Weinberger, Ali, and Rudd]{Kim1997}
Y.-K. Kim, W.~Hwang, N.~M. Weinberger, M.~A. Ali, and M.~E. Rudd.
\newblock {Electron-impact ionization cross sections of atmospheric molecules}.
\newblock \emph{J. Chem. Phys.}, 106\penalty0 (3):\penalty0 1026--1033, 01
  1997.
\newblock ISSN 0021-9606.
\newblock \doi{10.1063/1.473186}.
\newblock URL \url{https://doi.org/10.1063/1.473186}.

\bibitem[Kim and Desclaux(2002)]{Kim2002}
Yong-Ki Kim and Jean-Paul Desclaux.
\newblock Ionization of carbon, nitrogen, and oxygen by electron impact.
\newblock \emph{Phys. Rev. A}, 66:\penalty0 012708, July 2002.
\newblock \doi{10.1103/PhysRevA.66.012708}.
\newblock URL \url{https://link.aps.org/doi/10.1103/PhysRevA.66.012708}.

\bibitem[Kim and Rudd(1994)]{Kim1994}
Yong-Ki Kim and M.~Eugene Rudd.
\newblock Binary-encounter-dipole model for electron-impact ionization.
\newblock \emph{Phys. Rev. A}, 50:\penalty0 3954--3967, November 1994.
\newblock \doi{10.1103/PhysRevA.50.3954}.
\newblock URL \url{https://link.aps.org/doi/10.1103/PhysRevA.50.3954}.

\bibitem[Kim et~al.(2000)Kim, Santos, and Parente]{Kim2000}
Yong-Ki Kim, Jos\'e~Paulo Santos, and Fernando Parente.
\newblock Extension of the binary-encounter-dipole model to relativistic
  incident electrons.
\newblock \emph{Phys. Rev. A}, 62:\penalty0 052710, October 2000.
\newblock \doi{10.1103/PhysRevA.62.052710}.
\newblock URL \url{https://link.aps.org/doi/10.1103/PhysRevA.62.052710}.

\bibitem[Koga et~al.(1999)Koga, Kanayama, Watanabe, and Thakkar]{Koga1999}
Toshikatsu Koga, Katsutoshi Kanayama, Shinya Watanabe, and Ajit~J. Thakkar.
\newblock Analytical hartree–fock wave functions subject to cusp and
  asymptotic constraints: {He to Xe, Li$^+$ to Cs$^+$, H$^-$ to I$^-$}.
\newblock \emph{Int J Quantum Chem}, 71\penalty0 (6):\penalty0 491--497, 1999.
\newblock \doi{10.1002/(SICI)1097-461X(1999)71:6<491::AID-QUA6>3.0.CO;2-T}.
\newblock URL
  \url{https://doi.org/10.1002/(SICI)1097-461X(1999)71:6<491::AID-QUA6>3.0.CO;2-T}.

\bibitem[Koopmans(1934)]{Koopmans1934}
T~Koopmans.
\newblock {Über die Zuordnung von Wellenfunktionen und Eigenwerten zu den
  Einzelnen Elektronen Eines Atoms}.
\newblock \emph{Physica}, 1\penalty0 (1):\penalty0 104--113, 1934.
\newblock ISSN 0031-8914.
\newblock \doi{https://doi.org/10.1016/S0031-8914(34)90011-2}.
\newblock URL
  \url{https://www.sciencedirect.com/science/article/pii/S0031891434900112}.

\bibitem[Kramida et~al.(2022)Kramida, Ralchenko, Reader, and {NIST ASD
  Team}]{NIST-ADS}
A.~Kramida, Yu. Ralchenko, J.~Reader, and {NIST ASD Team}.
\newblock {NIST Atomic Spectra Database (version 5.9)}.
\newblock [Online], 2022.
\newblock URL \url{https://physics.nist.gov/asd}.
\newblock [Sat May 28 2022].

\bibitem[Kunhardt and Tzeng(1986)]{Kunhardt1986}
E.~E. Kunhardt and Y.~Tzeng.
\newblock Role of electron-molecule angular scattering in shaping the
  electron-velocity distribution.
\newblock \emph{Phys. Rev. A}, 34:\penalty0 2158--2166, September 1986.
\newblock \doi{10.1103/PhysRevA.34.2158}.
\newblock URL \url{https://link.aps.org/doi/10.1103/PhysRevA.34.2158}.

\bibitem[Köhn and Ebert(2014)]{Koehn2014}
C.~Köhn and U.~Ebert.
\newblock The structure of ionization showers in air generated by electrons
  with 1 {MeV} energy or less.
\newblock \emph{Plasma Sources Sci. Technol.}, 23\penalty0 (4):\penalty0
  045001, June 2014.
\newblock \doi{10.1088/0963-0252/23/4/045001}.
\newblock URL \url{https://doi.org/10.1088%2F0963-0252%2F23%2F4%2F045001}.

\bibitem[Lefebvre-Brion(1971)]{LefebvreBrion1971}
H.~Lefebvre-Brion.
\newblock Intensity anomaly in the photoelectron spectrum of {NO}.
\newblock \emph{Chem. Phys. Lett.}, 9\penalty0 (5):\penalty0 463--464, 1971.
\newblock ISSN 0009-2614.
\newblock \doi{https://doi.org/10.1016/0009-2614(71)80269-5}.
\newblock URL
  \url{https://www.sciencedirect.com/science/article/pii/0009261471802695}.

\bibitem[Lindsay and Mangan(2003)]{phe-Molc-5.1}
B.~G. Lindsay and M.~A. Mangan.
\newblock {5.1 Ionization}.
\newblock In Y.~Itikawa, editor, \emph{Interactions of Photons and Electrons
  with Molecules}, volume 17 C of \emph{Landolt-B{\"o}rnstein - Group I
  Elementary Particles, Nuclei and Atoms}. Springer-Verlag Berlin Heidelberg,
  2003.
\newblock \doi{10.1007/10874891_2}.
\newblock URL
  \url{https://materials.springer.com/lb/docs/sm_lbs_978-3-540-45843-2_2}.

\bibitem[Lindsay et~al.(2000)Lindsay, Mangan, Straub, and
  Stebbings]{Lindsay2000}
B.~G. Lindsay, M.~A. Mangan, H.~C. Straub, and R.~F. Stebbings.
\newblock {Absolute partial cross sections for electron-impact ionization of NO
  and NO2 from threshold to 1000 eV}.
\newblock \emph{J. Chem. Phys.}, 112\penalty0 (21):\penalty0 9404--9410, 06
  2000.
\newblock ISSN 0021-9606.
\newblock \doi{10.1063/1.481559}.
\newblock URL \url{https://doi.org/10.1063/1.481559}.

\bibitem[Liu et~al.(2000)Liu, An, Tang, Luo, Peng, and Long]{Liu2000}
Mantian Liu, Zhu An, Changhuan Tang, Zhengming Luo, Xiufeng Peng, and Xianguan
  Long.
\newblock Experimental electron-impact {K}-shell ionization cross sections.
\newblock \emph{Atom. Data Nucl. Data}, 76\penalty0 (2):\penalty0 213--234,
  2000.
\newblock ISSN 0092-640X.
\newblock \doi{https://doi.org/10.1006/adnd.2000.0843}.
\newblock URL
  \url{https://www.sciencedirect.com/science/article/pii/S0092640X0090843X}.

\bibitem[Llovet et~al.(2014)Llovet, Powell, Salvat, and Jablonski]{Llovet2014}
Xavier Llovet, Cedric~J. Powell, Francesc Salvat, and Aleksander Jablonski.
\newblock Cross sections for inner-shell ionization by electron impact.
\newblock \emph{J Phys Chem Ref Data}, 43\penalty0 (1):\penalty0 013102, 2014.
\newblock \doi{10.1063/1.4832851}.
\newblock URL \url{https://doi.org/10.1063/1.4832851}.

\bibitem[Lorquet and Desouter(1972)]{Lorquet1972}
J.C. Lorquet and M.~Desouter.
\newblock {Excited states of gaseous ions. Transitions to and predissociation
  of the C$\,{}^{2}\Sigma^+_u$ state of N$^+_2$}.
\newblock \emph{Chem Phys Lett}, 16\penalty0 (1):\penalty0 136--140, 1972.
\newblock ISSN 0009-2614.
\newblock \doi{https://doi.org/10.1016/0009-2614(72)80475-5}.
\newblock URL
  \url{https://www.sciencedirect.com/science/article/pii/0009261472804755}.

\bibitem[Lotz(1967)]{Lotz1967}
Wolfgang Lotz.
\newblock An empirical formula for the electron-impact ionization
  cross-section.
\newblock \emph{Z. Für Phys.}, 206\penalty0 (2):\penalty0 205--211, 1967.
\newblock ISSN 0044-3328.
\newblock \doi{10.1007/BF01325928}.
\newblock URL \url{https://doi.org/10.1007/BF01325928}.

\bibitem[Masuoka and Kobayashi(2004)]{Masuoka2004}
Toshio Masuoka and Ataru Kobayashi.
\newblock Single- and double-photoionization cross-sections of nitrogen dioxide
  {(NO$_2$)} and ionic fragmentation of {NO$_2^+$}and {NO$_2^{2+}$}.
\newblock \emph{Chem. Phys.}, 302\penalty0 (1):\penalty0 31--41, 2004.
\newblock ISSN 0301-0104.
\newblock \doi{https://doi.org/10.1016/j.chemphys.2004.03.009}.
\newblock URL
  \url{https://www.sciencedirect.com/science/article/pii/S0301010404001193}.

\bibitem[Mocellin et~al.(2001)Mocellin, Wiesner, Burmeister, Björneholm, and
  Naves~de Brito]{Mocellin2001}
A.~Mocellin, K.~Wiesner, F.~Burmeister, O.~Björneholm, and A.~Naves~de Brito.
\newblock {Experimental study of photoionization of ozone in the 12 to 21 eV
  region}.
\newblock \emph{J. Chem. Phys.}, 115\penalty0 (11):\penalty0 5041--5046, 09
  2001.
\newblock ISSN 0021-9606.
\newblock \doi{10.1063/1.1394945}.
\newblock URL \url{https://doi.org/10.1063/1.1394945}.

\bibitem[Newson et~al.(1995)Newson, Luc, Price, and Mason]{Newson1995}
Karl~A. Newson, Stephanie~M. Luc, Stephen~D. Price, and Nigel~J. Mason.
\newblock Electron-impact ionization of ozone.
\newblock \emph{Int. J. Mass Spectrom.}, 148\penalty0 (3):\penalty0 203--213,
  1995.
\newblock ISSN 0168-1176.
\newblock \doi{https://doi.org/10.1016/0168-1176(95)04300-A}.
\newblock URL
  \url{https://www.sciencedirect.com/science/article/pii/016811769504300A}.

\bibitem[Orient and Srivastava(1987)]{Orient1987}
O~J Orient and S~K Srivastava.
\newblock {Electron impact ionisation of H2O, CO, CO2 and CH4}.
\newblock \emph{J. Phys. B: At. Mol. Phys.}, 20\penalty0 (15):\penalty0 3923,
  aug 1987.
\newblock \doi{10.1088/0022-3700/20/15/036}.
\newblock URL \url{https://dx.doi.org/10.1088/0022-3700/20/15/036}.

\bibitem[Perkins et~al.(1991)Perkins, Cullen, Chen, Rathkopf, Scofield, and
  Hubbell]{EADL}
S.~T. Perkins, D.~E. Cullen, M.~H. Chen, J.~Rathkopf, J.~Scofield, and J.~H.
  Hubbell.
\newblock Tables and graphs of atomic subshell and relaxation data derived from
  the {LLNL Evaluated Atomic Data Library (EADL), Z = 1--100}.
\newblock Technical report, Lawrence Livermore National Lab. (LLNL), 10 1991.
\newblock URL \url{https://www.osti.gov/biblio/10121422}.

\bibitem[Price(1981)]{Price1981}
W.~C. Price.
\newblock Photoelectron spectroscopy and molecular structure.
\newblock \emph{Int. Rev. Phys. Chem.}, 1\penalty0 (1):\penalty0 1--30, 1981.
\newblock \doi{10.1080/01442358109353240}.
\newblock URL \url{https://doi.org/10.1080/01442358109353240}.

\bibitem[Rapp and Englander‐Golden(1965)]{Rapp1965}
Donald Rapp and Paula Englander‐Golden.
\newblock Total cross sections for ionization and attachment in gases by
  electron impact. {I.} {P}ositive ionization.
\newblock \emph{J. Chem. Phys.}, 43\penalty0 (5):\penalty0 1464--1479, 1965.
\newblock \doi{10.1063/1.1696957}.
\newblock URL \url{https://doi.org/10.1063/1.1696957}.

\bibitem[Rejoub et~al.(2002)Rejoub, Lindsay, and Stebbings]{Rejoub2002}
R.~Rejoub, B.~G. Lindsay, and R.~F. Stebbings.
\newblock Determination of the absolute partial and total cross sections for
  electron-impact ionization of the rare gases.
\newblock \emph{Phys. Rev. A}, 65:\penalty0 042713, Apr 2002.
\newblock \doi{10.1103/PhysRevA.65.042713}.
\newblock URL \url{https://link.aps.org/doi/10.1103/PhysRevA.65.042713}.

\bibitem[Ridenti et~al.(2015)Ridenti, Alves, Guerra, and Amorim]{Ridenti2015}
M.~A. Ridenti, L.~L. Alves, V.~Guerra, and J.~Amorim.
\newblock The role of rotational mechanisms in electron swarm parameters at low
  reduced electric field in {N$_2$}, {O$_2$}and {H$_2$}.
\newblock \emph{Plasma Sources Sci. Technol.}, 24\penalty0 (3):\penalty0
  035002, April 2015.
\newblock \doi{10.1088/0963-0252/24/3/035002}.
\newblock URL \url{https://doi.org/10.1088/0963-0252/24/3/035002}.

\bibitem[Sant'Anna et~al.(2011)Sant'Anna, Schlachter, \"Ohrwall, Stolte,
  Lindle, and McLaughlin]{SantAnna2011}
M.~M. Sant'Anna, A.~S. Schlachter, G.~\"Ohrwall, W.~C. Stolte, D.~W. Lindle,
  and B.~M. McLaughlin.
\newblock {$K$}-shell {X}-ray spectroscopy of atomic nitrogen.
\newblock \emph{Phys. Rev. Lett.}, 107:\penalty0 033001, Jul 2011.
\newblock \doi{10.1103/PhysRevLett.107.033001}.
\newblock URL \url{https://link.aps.org/doi/10.1103/PhysRevLett.107.033001}.

\bibitem[Santos et~al.(2003)Santos, Parente, and Kim]{Santos2003}
J.~P. Santos, F.~Parente, and Y.-K. Kim.
\newblock Cross sections for {K}-shell ionization of atoms by electron impact.
\newblock \emph{J. Phys. B: At. Mol. Opt. Phys.}, 36\penalty0 (21):\penalty0
  4211--4224, October 2003.
\newblock \doi{10.1088/0953-4075/36/21/002}.
\newblock URL \url{https://doi.org/10.1088%2F0953-4075%2F36%2F21%2F002}.

\bibitem[Schmalzried(2023)]{Schmalzried2023}
A.~Schmalzried.
\newblock \emph{Electron thermal runaway in atmospheric electrified gases: a
  microscopic approach}.
\newblock PhD thesis, Instituto de Astrofísica de Andalucía, 2023.
\newblock URL \url{http://hdl.handle.net/10261/338079}.

\bibitem[Schram et~al.(1965)Schram, {De Heer}, {van der Wiel}, and
  Kistemaker]{Schram1965}
B.~L. Schram, F.~J. {De Heer}, M.~J. {van der Wiel}, and J.~Kistemaker.
\newblock Ionization cross sections for electrons (0.6–20 {keV}) in noble and
  diatomic gases.
\newblock \emph{Physica}, 31\penalty0 (1):\penalty0 94--112, 1965.
\newblock ISSN 0031-8914.
\newblock \doi{https://doi.org/10.1016/0031-8914(65)90109-6}.
\newblock URL
  \url{https://www.sciencedirect.com/science/article/pii/0031891465901096}.

\bibitem[Schram et~al.(1966)Schram, Moustafa, Schutten, and {de
  Heer}]{Schram1966}
B.~L. Schram, H.~R. Moustafa, J.~Schutten, and F.~J. {de Heer}.
\newblock Ionization cross sections for electrons (100–600 {eV}) in noble and
  diatomic gases.
\newblock \emph{Physica}, 32\penalty0 (4):\penalty0 734--740, 1966.
\newblock ISSN 0031-8914.
\newblock \doi{https://doi.org/10.1016/0031-8914(66)90005-X}.
\newblock URL
  \url{https://www.sciencedirect.com/science/article/pii/003189146690005X}.

\bibitem[Shen et~al.(2018)Shen, Wang, Gong, Shan, and Chen]{Shen2018}
Zhenjie Shen, Enliang Wang, Maomao Gong, Xu~Shan, and Xiangjun Chen.
\newblock Electron-impact ionization cross sections for nitrogen molecule from
  250 to 8000 {eV}.
\newblock \emph{J Electron Spectrosc}, 225:\penalty0 42--48, 2018.
\newblock ISSN 0368-2048.
\newblock \doi{https://doi.org/10.1016/j.elspec.2018.03.009}.
\newblock URL
  \url{http://www.sciencedirect.com/science/article/pii/S0368204817302931}.

\bibitem[Shibuya et~al.(1997)Shibuya, Suzuki, Imamura, and Koyano]{Shibuya1997}
Kazuhiko Shibuya, Shinzo Suzuki, Takashi Imamura, and Inosuke Koyano.
\newblock Dissociation of state-selected {NO2+} ions studied by threshold
  photoelectron--photoion coincidence techniques.
\newblock \emph{J. Phys. Chem. A}, 101\penalty0 (4):\penalty0 685--689, January
  1997.
\newblock ISSN 1089-5639.
\newblock \doi{10.1021/jp962031f}.
\newblock URL \url{https://doi.org/10.1021/jp962031f}.

\bibitem[Shimamura and Takayanagi(1984)]{eMolColl}
I.~Shimamura and K.~Takayanagi, editors.
\newblock \emph{Electron-molecule collisions}.
\newblock Physics of atoms and molecules. Plenum Press, New York [etc.], 1984.
\newblock ISBN 0-306-41531-3.
\newblock \doi{10.1007/978-1-4613-2357-0}.
\newblock URL \url{https://link.springer.com/book/10.1007/978-1-4613-2357-0}.

\bibitem[Siegbahn et~al.(1969)Siegbahn, Nordling, Johansson, Hedman, Hed{\'e}n,
  Hamrin, Gelius, Bergmark, Werme, Manne, et~al.]{ESCA2}
K.~Siegbahn, C.~Nordling, G.~Johansson, J.~Hedman, P.~F. Hed{\'e}n, K.~Hamrin,
  U.~Gelius, T.~Bergmark, L.~O. Werme, R.~Manne, et~al.
\newblock \emph{{ESCA: Applied to Free Molecules}}.
\newblock North-Holland Publishing Company, 1969.
\newblock URL \url{https://books.google.es/books?id=PM_bxQEACAAJ}.

\bibitem[Siegbahn et~al.(1967)Siegbahn, Nordling, and Fahlman]{ESCA1}
Kai Siegbahn, Carl Nordling, and Anders Fahlman.
\newblock \emph{{ESCA} : atomic, molecular and solid state structure studied by
  means of electron spectroscopy}.
\newblock Almqvist and Wiksell, Uppsala, 1967.
\newblock URL \url{http://lib.ugent.be/catalog/rug01:001343982}.

\bibitem[{\v{S}}imek and Bonaventura(2018)]{Simek2018}
M.~{\v{S}}imek and Z.~Bonaventura.
\newblock Non-equilibrium kinetics of the ground and excited states in
  {N$_2$}{\textendash}{O$_2$} under nanosecond discharge conditions: extended
  scheme and comparison with available experimental observations.
\newblock \emph{J. Phys. D: Appl. Phys.}, 51\penalty0 (50):\penalty0 504004,
  October 2018.
\newblock \doi{10.1088/1361-6463/aadcd1}.
\newblock URL \url{https://doi.org/10.1088/1361-6463/aadcd1}.

\bibitem[Straub et~al.(1996)Straub, Renault, Lindsay, Smith, and
  Stebbings]{Straub1996}
H.~C. Straub, P.~Renault, B.~G. Lindsay, K.~A. Smith, and R.~F. Stebbings.
\newblock Absolute partial cross sections for electron-impact ionization of
  {H$_2$, N$_2$, and O$_2$} from threshold to 1000 {eV}.
\newblock \emph{Phys. Rev. A}, 54:\penalty0 2146--2153, September 1996.
\newblock \doi{10.1103/PhysRevA.54.2146}.
\newblock URL \url{https://link.aps.org/doi/10.1103/PhysRevA.54.2146}.

\bibitem[Straub et~al.(1998)Straub, Lindsay, Smith, and Stebbings]{Straub1998}
H.~C. Straub, B.~G. Lindsay, K.~A. Smith, and R.~F. Stebbings.
\newblock {Absolute partial cross sections for electron-impact ionization of
  H2O and D2O from threshold to 1000 eV}.
\newblock \emph{J. Chem. Phys.}, 108\penalty0 (1):\penalty0 109--116, 01 1998.
\newblock ISSN 0021-9606.
\newblock \doi{10.1063/1.475367}.
\newblock URL \url{https://doi.org/10.1063/1.475367}.

\bibitem[Tan et~al.(1978)Tan, Brion, {Van der Leeuw}, and {van der
  Wiel}]{Tan1978}
K.H. Tan, C.E. Brion, Ph.E. {Van der Leeuw}, and M.J. {van der Wiel}.
\newblock Absolute oscillator strengths {(10–60 eV)} for the photoabsorption,
  photoionisation and fragmentation of {H20s}.
\newblock \emph{Chem. Phys.}, 29\penalty0 (3):\penalty0 299--309, 1978.
\newblock ISSN 0301-0104.
\newblock \doi{https://doi.org/10.1016/0301-0104(78)85080-0}.
\newblock URL
  \url{https://www.sciencedirect.com/science/article/pii/0301010478850800}.

\bibitem[Tanaka et~al.(2016)Tanaka, Brunger, Campbell, Kato, Hoshino, and
  Rau]{Tanaka2016}
H.~Tanaka, M.~J. Brunger, L.~Campbell, H.~Kato, M.~Hoshino, and A.~R.~P. Rau.
\newblock Scaled plane-wave {B}orn cross sections for atoms and molecules.
\newblock \emph{Rev. Mod. Phys.}, 88:\penalty0 025004, May 2016.
\newblock \doi{10.1103/RevModPhys.88.025004}.
\newblock URL \url{https://link.aps.org/doi/10.1103/RevModPhys.88.025004}.

\bibitem[Tawara et~al.(1973)Tawara, Harrison, and {De Heer}]{Tawara1973}
H.~Tawara, K.G. Harrison, and F.J. {De Heer}.
\newblock X-ray emission cross sections and fluorescence yields for light atoms
  and molecules by electron impact.
\newblock \emph{Physica}, 63\penalty0 (2):\penalty0 351--367, 1973.
\newblock ISSN 0031-8914.
\newblock \doi{https://doi.org/10.1016/0031-8914(73)90321-2}.
\newblock URL
  \url{https://www.sciencedirect.com/science/article/pii/0031891473903212}.

\bibitem[Tayal and Zatsarinny(2016)]{Tayal2016O}
S.~S. Tayal and Oleg Zatsarinny.
\newblock {$B$-spline $R$-matrix-with-pseudostates} approach for excitation and
  ionization of atomic oxygen by electron collisions.
\newblock \emph{Phys. Rev. A}, 94:\penalty0 042707, October 2016.
\newblock \doi{10.1103/PhysRevA.94.042707}.
\newblock URL \url{https://link.aps.org/doi/10.1103/PhysRevA.94.042707}.

\bibitem[Thompson et~al.(1995)Thompson, Shah, and Gilbody]{Thompson1995}
W~R Thompson, M~B Shah, and H~B Gilbody.
\newblock Single and double ionization of atomic oxygen by electron impact.
\newblock \emph{J. Phys. B: At. Mol. Opt. Phys.}, 28\penalty0 (7):\penalty0
  1321--1330, April 1995.
\newblock \doi{10.1088/0953-4075/28/7/023}.
\newblock URL \url{https://doi.org/10.1088/0953-4075/28/7/023}.

\bibitem[Vriens(1966)]{Vriens1966}
L~Vriens.
\newblock Electron exchange in binary encounter collision theory.
\newblock \emph{Proc. Phys. Soc.}, 89\penalty0 (1):\penalty0 13--21, September
  1966.
\newblock \doi{10.1088/0370-1328/89/1/304}.
\newblock URL \url{https://doi.org/10.1088/0370-1328/89/1/304}.

\bibitem[Wang et~al.(2013)Wang, Zatsarinny, and Bartschat]{Wang2013C}
Yang Wang, Oleg Zatsarinny, and Klaus Bartschat.
\newblock $b$-spline $r$-matrix-with-pseudostates calculations for
  electron-impact excitation and ionization of carbon.
\newblock \emph{Phys. Rev. A}, 87:\penalty0 012704, Jan 2013.
\newblock \doi{10.1103/PhysRevA.87.012704}.
\newblock URL \url{https://link.aps.org/doi/10.1103/PhysRevA.87.012704}.

\bibitem[Wang et~al.(2014)Wang, Zatsarinny, and Bartschat]{Wang2014N}
Yang Wang, Oleg Zatsarinny, and Klaus Bartschat.
\newblock {B}-spline {R}-matrix-with-pseudostates calculations for
  electron-impact excitation and ionization of nitrogen.
\newblock \emph{Phys. Rev. A}, 89:\penalty0 062714, June 2014.
\newblock \doi{10.1103/PhysRevA.89.062714}.
\newblock URL \url{https://link.aps.org/doi/10.1103/PhysRevA.89.062714}.

\bibitem[Wiesner et~al.(2003)Wiesner, Fink, Sorensen, Andersson, Feifel,
  Hjelte, Miron, {Naves de Brito}, Rosenqvist, Wang, Svensson, and
  Björneholm]{Wiesner2003}
K~Wiesner, R.F Fink, S.L Sorensen, M~Andersson, R~Feifel, I~Hjelte, C~Miron,
  A~{Naves de Brito}, L~Rosenqvist, H~Wang, S~Svensson, and O~Björneholm.
\newblock Valence photoionization and resonant core excitation of ozone –
  experimental and theoretical study of the {$\tilde{\text{C}}$-state of O3+}.
\newblock \emph{Chem. Phys. Lett.}, 375\penalty0 (1):\penalty0 76--83, 2003.
\newblock ISSN 0009-2614.
\newblock \doi{https://doi.org/10.1016/S0009-2614(03)00818-2}.
\newblock URL
  \url{https://www.sciencedirect.com/science/article/pii/S0009261403008182}.

\bibitem[Younger and M{\"a}rk(1985)]{EII-2}
S.~M. Younger and T.~D. M{\"a}rk.
\newblock \emph{Semi-Empirical and Semi-Classical Approximations for Electron
  Ionization}, chapter~2, pages 24--41.
\newblock Springer Vienna, Vienna, 1985.
\newblock ISBN 978-3-7091-4028-4.
\newblock \doi{10.1007/978-3-7091-4028-4_2}.
\newblock URL \url{https://doi.org/10.1007/978-3-7091-4028-4_2}.

\bibitem[Zatsarinny and Bartschat(2013)]{Zatsarinny2013}
Oleg Zatsarinny and Klaus Bartschat.
\newblock {The B-spline R-matrix method for atomic processes: application to
  atomic structure, electron collisions and photoionization}.
\newblock \emph{J. Phys. B: At. Mol. Opt. Phys.}, 46\penalty0 (11):\penalty0
  112001, May 2013.
\newblock \doi{10.1088/0953-4075/46/11/112001}.
\newblock URL \url{https://dx.doi.org/10.1088/0953-4075/46/11/112001}.

\end{thebibliography}

\clearpage

\end{document}